\documentclass[]{spie}  
\usepackage[]{graphicx}
\usepackage{amsmath,amssymb,mathtools,url}
\title{\Large \bf Vortex fiber nulling for exoplanet observations: conceptual design, theoretical performance, and initial scientific yield predictions }
\usepackage{array}
\usepackage{hyperref}
\newcolumntype{P}[1]{>{\centering\arraybackslash}p{#1}}

\author{
Garreth~Ruane\supit{a,b}, Daniel~Echeverri\supit{b}, Nemanja~Jovanovic\supit{b}, Dimitri~Mawet\supit{b,a}, Eugene~Serabyn\supit{a} J.~Kent~Wallace\supit{a}, Jason~Wang\supit{b}, and Natasha~Batalha\supit{c}\\
\supit{a}Jet Propulsion Laboratory, California Institute of Technology, 4800 Oak Grove Dr., \\Pasadena, CA 91109, USA\\
\supit{b}Department of Astronomy, California Institute of Technology, 1200 E. California Blvd., \\Pasadena, CA 91125, USA \\
\supit{c}Department of Astronomy and Astrophysics, University of California Santa Cruz, \\Santa Cruz, CA 95064, USA 
}

\authorinfo{$^\dagger$Send correspondence to gruane@jpl.nasa.gov}

\graphicspath{{Figures/}}
 
\begin{document} 
  \maketitle 

\begin{abstract}
Vortex fiber nulling (VFN) is a method that may enable the detection and characterization of exoplanets at small angular separations (0.5-2~$\lambda/D$) with ground- and space-based telescopes. Since the field of view is within the inner working angle of most coronagraphs, nulling accesses non-transiting planets that are otherwise too close to their star for spectral characterization by other means, thereby significantly increasing the number of known exoplanets available for direct spectroscopy in the near-infrared. Furthermore, VFN targets planets on closer-in orbits which tend to have more favorable planet-to-star flux ratios in reflected light. Here, we present the theory and applications of VFN, show that the optical performance is approximately equivalent for a variety of implementations and aperture shapes, and discuss the trade-offs between throughput and engineering requirements using numerical simulations. We compare vector and scalar approaches and, finally, show that beam shaping optics may be used to significantly improve the throughput for planet light. Based on theoretical performance, we estimate the number of known planets and theoretical exoEarths accessible with a VFN instrument linked to a high-resolution spectrograph on the future Thirty Meter Telescope. 
\end{abstract}



\section{INTRODUCTION}
\label{sec:intro} 

Nulling interferometry is used to reduce the intensity of starlight prior to detection while allowing for the detection and spectral characterization of exoplanets with acceptable losses in planet signal, resulting in a significant improvement to the raw signal-to-noise ratio ($S/N$). Ronald N. Bracewell originally proposed using an interferometer to null starlight for the purpose of exoplanet detection in 1978\cite{Bracewell1978}. Bracewell's interferometer used two symmetric apertures to create an elongated interference pattern, or fringes, which could be rotated in time to modulate the planet signal, while the stellar leakage remained constant\cite{Bracewell1979}. Building on this concept, Haguenauer and Serabyn (2006)\cite{Haguenauer2006} showed that a single-mode optical fiber placed in the image plane improved the rejection of starlight in the Bracewell configuration. An alternate interferometric concept proposed by Swartzlander (2001)\cite{Swartzlander2001} used a phase mask in the pupil plane, rather than individual sub-apertures, to create an optical vortex in the image plane with a central null and a rotationally-symmetric bright fringe. Vortex fiber nulling (VFN)\cite{Ruane2018_VFN, Echeverri2019_VFN} combines elements from all of these concepts to form a simple, single-aperture, broadband fiber nulling interferometer with rotationally-symmetric throughput for planet light. 

In scenarios where an exoplanet is barely resolved from its host star, VFN selectively suppresses the starlight while light from the planet may be efficiently routed to a diffraction-limited spectrograph via a single-mode optical fiber. The improved $S/N$ afforded by VFN may enable astronomers to detect the planet and potentially characterize its atmosphere by reducing the integration time required to do so by orders of magnitude. Although the optimal spectral resolution for this purpose remains an open question, with sufficiently high spectral resolution, cross-correlation techniques offer the potential to characterize the atmosphere of exoplanets in reflected light with modest $S/N$ per spectral channel\cite{SparksFord2002,Riaud2007,Snellen2015,Wang2017}. When combined with such techniques, VFN is well suited for follow up observations of planets that were previously detected by radial velocity or transit methods, such as the Earth-sized planets in the habitable zone of nearby M~stars: Proxima~Centauri~b\cite{ProxCen} and Ross~128~b\cite{Bonfils2017}.

The two key optical components for VFN are a vortex phase mask and a single-mode fiber (SMF). Vortex phase masks are also utilized in vortex coronagraphs\cite{Mawet2005,Foo2005}, which are currently operating on several ground-based telescopes\cite{Mawet2010b,Serabyn2010,Serabyn2017,Mawet2017,Ruane2017,Ruane2019_RDI} and are a leading design for future space telescopes with coronagraphs\cite{Ruane2018_JATIS}. VFN-capable instruments will take advantage of more than a decade of technological development to improve the quality and bandwidth of vortex phase masks at visible and infrared wavelengths for the purpose of high-contrast imaging\cite{Delacroix2013,SerabynTDEM1,SerabynTDEM2,Serabyn2019}. Coronagraph instruments are also starting to take advantage of SMFs for improving the rejection of stellar speckles in cases where the planet is spatially resolved from the star\cite{Mawet2017_HDCII,Llop2019}. Thus, the optics required for VFN are readily available and may be commonplace in future adaptive optics (AO) instruments designed for exoplanet science. 

While multi-aperture fiber nulling has been demonstrated on sky using the 200-inch Hale Telescope at Palomar Observatory\cite{Serabyn2019_PFN} and preparation is underway for the first on-sky demonstration of VFN as part of the Keck Planet Imager and Characterizer (KPIC) instrument\cite{Mawet2016_KPIC,Mawet2017_KPIC,Echeverri2019b_VFN}, in the following we present conceptual design trades, theoretical performance, and estimated scientific yield of a VFN mode operating in the near-infrared on the future Thirty Meter Telescope (TMT) Multi-Object Diffraction-limited High-resolution Infrared Spectrograph (MODHIS) instrument\cite{Mawet2019_whitepaper}. Specifically, we will show that it is feasible to detect $>$10 known planets with integration times of tens of hours and potentially search for rocky planets in the habitable zones of nearby stars. 

\begin{figure}[t!]
    \centering
    \includegraphics[width=\linewidth]{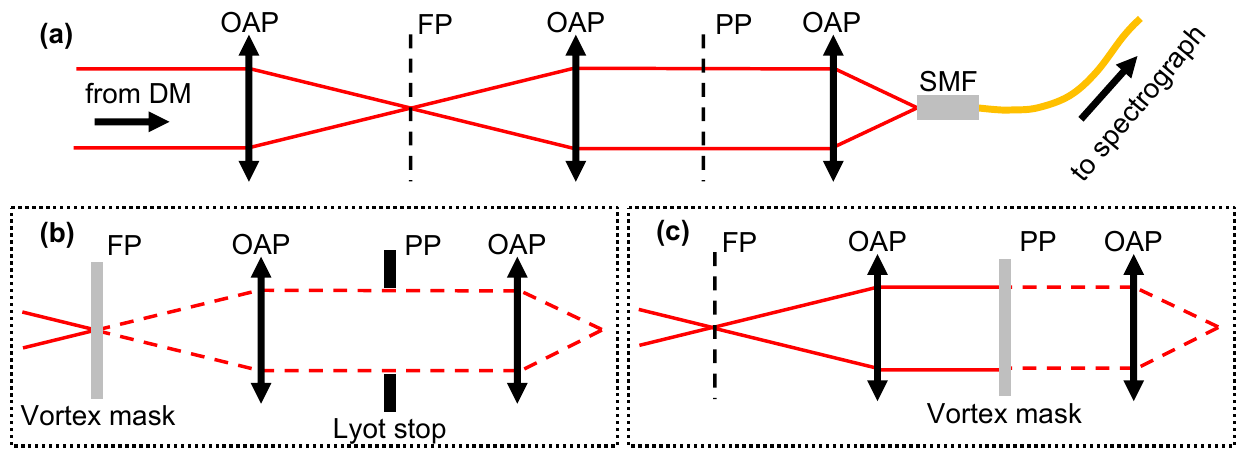}
    \caption{(a) A schematic optical layout for an AO-fed fiber injection unit for exoplanet spectroscopy. VFN may be implemented in a number of ways, including configurations with the vortex mask (b)~in a focal plane and (c)~in a pupil. DM:~Deformable mirror. OAP:~Off-axis parabola. FP:~Focal plane. PP:~Pupil plane. SMF:~Single-mode fiber. 
}
    \label{fig:opticallayouts}
\end{figure}

\section{VORTEX FIBER NULLING CONCEPT}

A vortex fiber nuller is formed by introducing a vortex mask within an AO-fed fiber injection unit (FIU), which forms a link between the telescope's AO system and a spectrograph. Figure \ref{fig:opticallayouts}a shows a representative optical layout of such an instrument. Light from astronomical sources couples into the SMF with coupling efficiency given by the square-magnitude of the so-called overlap integral:
\begin{equation}
    \eta=\left|\int E(\mathbf{r}) \Psi(\mathbf{r})dA\right|^2,
    \label{eqn:couplingeff}
\end{equation}
where $\mathbf{r}=(r,\theta)$ are the polar coordinates in the plane of the SMF entrance face, $E(\mathbf{r})$ is the complex field, and $\Psi(\mathbf{r})$ is the fiber mode profile\cite{Shaklan1988,Jeunhomme1989}. A fiber nuller\cite{Haguenauer2006,Martin2017} creates a condition where the stellar field is orthogonal to $\Psi(\mathbf{r})$ and thus does not couple into the SMF, while light from planets within a certain range of angular separations couples into the SMF with reasonable efficiency. 

\begin{figure}[t!]
    \centering
    \includegraphics[width=\linewidth]{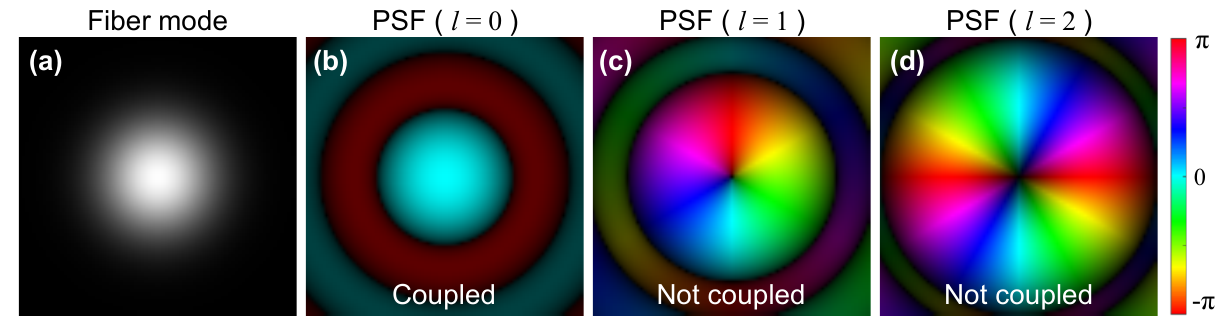}
    \caption{(a)~The fundamental mode of a single-mode step-index fiber. (b)~The nominal complex point spread function (PSF) of a telescope couples into a SMF, though with imperfect efficiency due to the mismatch between the mode and incident electric field profile. (c-d) However, a PSF containing an optical vortex at its center is orthogonal to the fiber mode and therefore does not couple into the SMF. (c) and (d) show cases with charges $l=1$ and $l=2$, respectively.}
    \label{fig:fibermodes}
\end{figure}

The principle of VFN is that the stellar beam is made into an optical vortex. For example, the field may take the form: $E(\mathbf{r})=f(r)\exp(il\theta)$, where $f(r)$ is the radial component of the field amplitude and $l$ is an integer known as the charge. The fundamental mode of a single-mode step-index fiber is also rotationally symmetric and approximately Gaussian. In this case, the overlap integral is separable and the polar term is  
\begin{equation}
    \int_0^{2\pi} \exp(il\theta)d\theta = 
    \left\{
    \begin{array}{ll}
          2\pi & l=0 \\
          0 & l\ne0 \\
    \end{array} 
    \right..
    \label{eqn:overlap_az}
\end{equation}
Thus, for $l\ne0$, the stellar field is orthogonal to the SMF mode and is therefore rejected by the SMF. Figure~\ref{fig:fibermodes} shows the fiber mode along with example stellar fields with $l=0$, which couples into the SMF, and with $l=1$ and $l=2$, which do not couple. While the rotationally symmetric amplitude profile is used here as a convenient example, Eqn.~\ref{eqn:couplingeff} computes to zero for a variety of symmetries for the complex fields produced in telescopes and SMF modes.  

In theory, the optical vortex can be introduced anywhere along the stellar beam path. Figs.~\ref{fig:opticallayouts}b,c show two examples where the phase would be singular along chief ray in an otherwise perfect system. In Fig.~\ref{fig:opticallayouts}b, the layout is similar to that of a vortex coronagraph\cite{Mawet2005,Foo2005}, which has the vortex mask in an intermediate focal plane and a so-called Lyot stop in the downstream pupil. On the other hand, Fig.~\ref{fig:opticallayouts}c has the vortex mask in a pupil plane. We will show in the next section that these arrangements perform similarly in terms of starlight suppression and the coupling of planet light.

\begin{figure}[t!]
    \centering
    \includegraphics[width=0.9\linewidth]{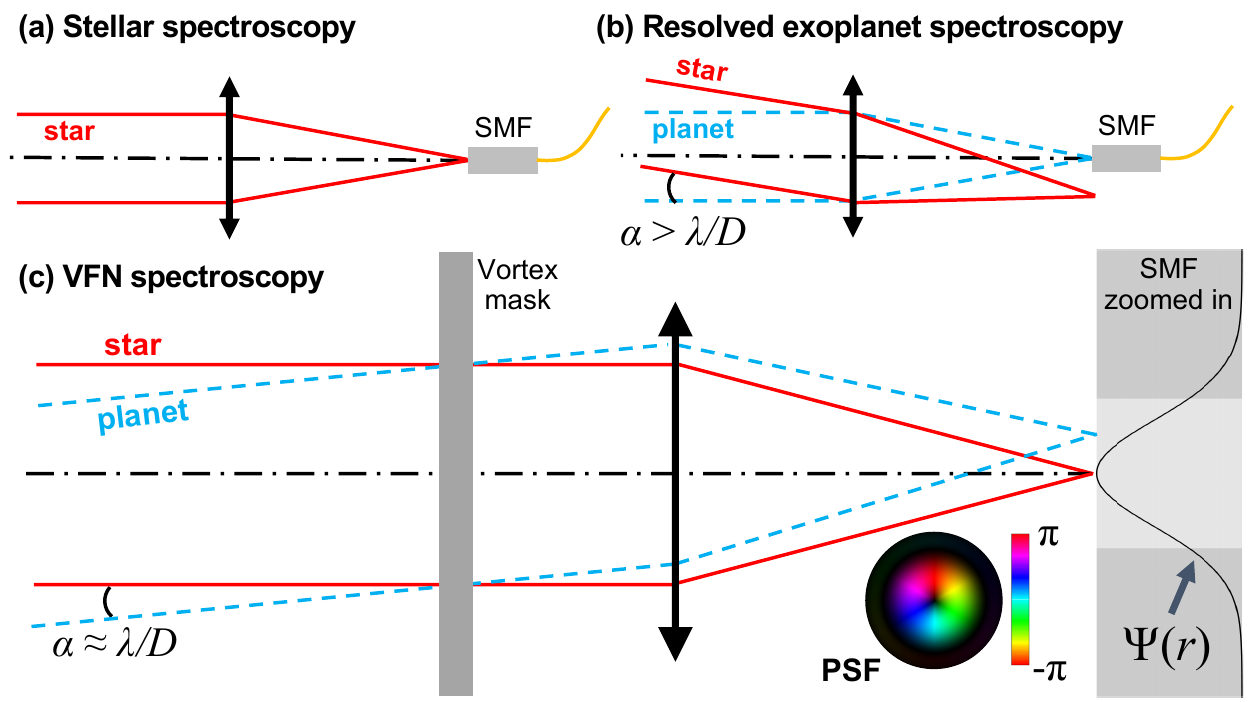}
    \caption{Observing configurations of an AO-fed FIU. (a)~For stellar spectroscopy, the star is coupled directly into the SMF. (b)~For resolved exoplanet spectroscopy, the planet is aligned to the optical axis and the stellar image appears at an angular separation $>\lambda/D$. (c)~VFN is used when the star and planet are separated by $\sim\lambda/D$. The stellar beam containing the optical vortex (see inset) is centered on the SMF and nulled, while the planet is slightly off-axis and partially couples into the SMF. The inset shows the phase of the complex field of the stellar PSF.}
    \label{fig:obs_scenarios}
\end{figure}

For context, an AO-fed FIU has several potential observing configurations (see Fig.~\ref{fig:obs_scenarios}). Most commonly, the starlight is coupled directly into the SMF with high efficiency for stellar spectroscopy (Fig.~\ref{fig:obs_scenarios}a). For resolved exoplanet spectroscopy (Fig.~\ref{fig:obs_scenarios}b), the planet is aligned to the optical axis and the star is imaged at an angular separation $>\lambda/D$, where $\lambda$ is the wavelength and $D$ is the telescope diameter. In this case, an apodizer or coronagraph may be introduced to reduce the diffracted starlight at the position of the planet. VFN is advantageous when the star and planet of interest are separated by approximately $\lambda/D$. As shown in Fig.~\ref{fig:obs_scenarios}c, the stellar beam containing the optical vortex is centered on the SMF and nulled, while the planet is slightly off-axis and therefore partially couples into the fiber.

\section{THEORETICAL PERFORMANCE AND OPTICAL DESIGN TRADES}

In this section, we present the simulated performance of example VFN instrument layouts on different telescopes, including their sensitivity to a variety of errors. We also discuss the pros and cons of pupil- and focal-plane VFN implementations as well as vector versus scalar vortex masks. Finally, we describe ways to use beam shaping optics in order to improve coupling efficiency. 

\subsection{Throughput}

We define the throughput of a VFN instrument as the fraction of light from a distant point source that enters the SMF, normalized by the total energy. In the following, we ignore imperfect reflectance/transmission of the focusing and collimating optics, Fresnel losses at the ends of the SMF, and propagation losses within the SMF. The throughput is computed using Eqn.~\ref{eqn:couplingeff} and depends on the angular separation of the point source from the optical axis, $\alpha$. The throughput and coupling efficiency are equal if there is no Lyot stop in the system. The term ``null depth" refers to the throughput for the star when the starlight is intentionally nulled. 

\begin{figure}
    \centering
    \includegraphics[width=0.75\linewidth]{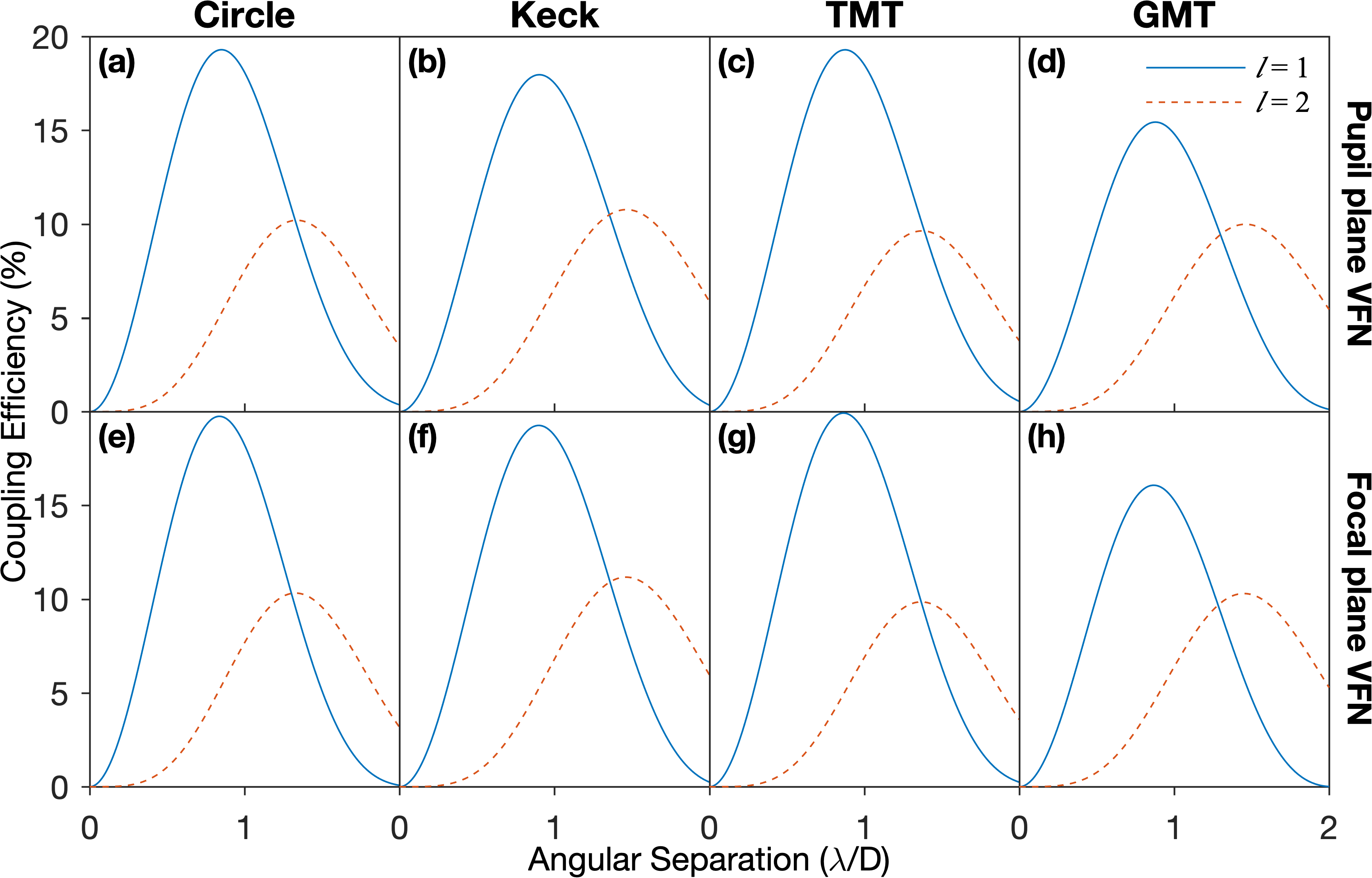}
    \caption{Optimal coupling efficiency, $\eta$, for a point source versus its angular separation from the optical axis, $\alpha$. These curves are calculated at the central wavelength assuming the optimal value of $q$ (see Table \ref{tab:optparams}). Results are shown with the vortex mask placed in (a)-(d)~the pupil and (e)-(h)~the focal plane. The columns correspond to the pupil shapes shown in Fig.~\ref{fig:pupils}.}
    \label{fig:throughput_grid}
\end{figure}

\begin{figure}[t]
    \centering
    \includegraphics[width=0.8\linewidth]{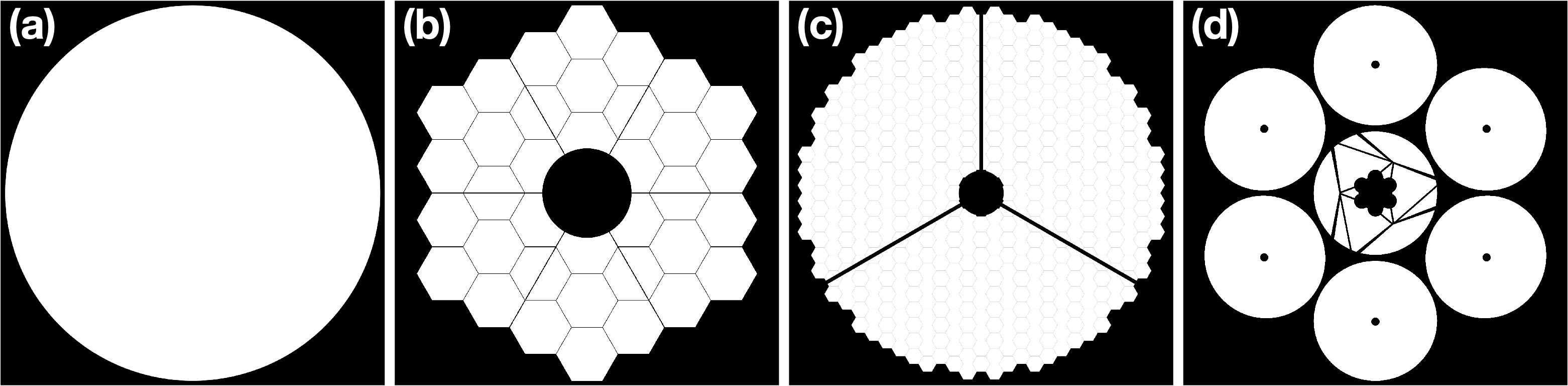}
    \caption{Pupil shapes for (a)~a circular aperture, (b)~the Keck Telescopes, (c)~the Thirty Meter Telescope (TMT), and (d)~the Giant Magellan Telescope (GMT). 
    }
    \label{fig:pupils}
\end{figure}

Figure~\ref{fig:throughput_grid} shows the throughput for four example aperture shapes, with focal- and pupil-plane VFN configurations (Figs.~\ref{fig:opticallayouts}b,c), as well as charge $l=1$ and $l=2$ vortex masks. The aperture shapes (see Fig.~\ref{fig:pupils}) correspond to a circular aperture, the Keck telescopes, the Thirty Meter Telescope (TMT), and the Giant Magellan Telescope (GMT). We have omitted the Lyot stop for the sake of simplicity. Each case in Fig.~\ref{fig:throughput_grid} confirms that the coupling efficiency, $\eta$, is zero where $\alpha=0$ (i.e. the starlight is nulled) and that off-axis sources partially couple into the fiber. The VFN performance is relatively insensitive to the shape of the aperture (Fig.~\ref{fig:throughput_grid}, columns) or whether the vortex mask is situated in the pupil or focal plane (Fig.~\ref{fig:throughput_grid}, rows), which differs considerably from the trade-offs typically encountered for coronagraphs\cite{Ruane2018_FALCO}. The maximum coupling efficiency, $\eta_\text{peak}$, is approximately 20\% for $l=1$ and 10\% for $\l=2$. The peaks occur at $\alpha_\mathrm{peak}=0.9~\lambda/D$ and $\alpha_\mathrm{peak}=1.3~\lambda/D$, respectively. See Table~\ref{tab:optparams} for a more complete list of important parameters for each case. 

\begin{figure}
    \centering
    \includegraphics[width=0.8\linewidth]{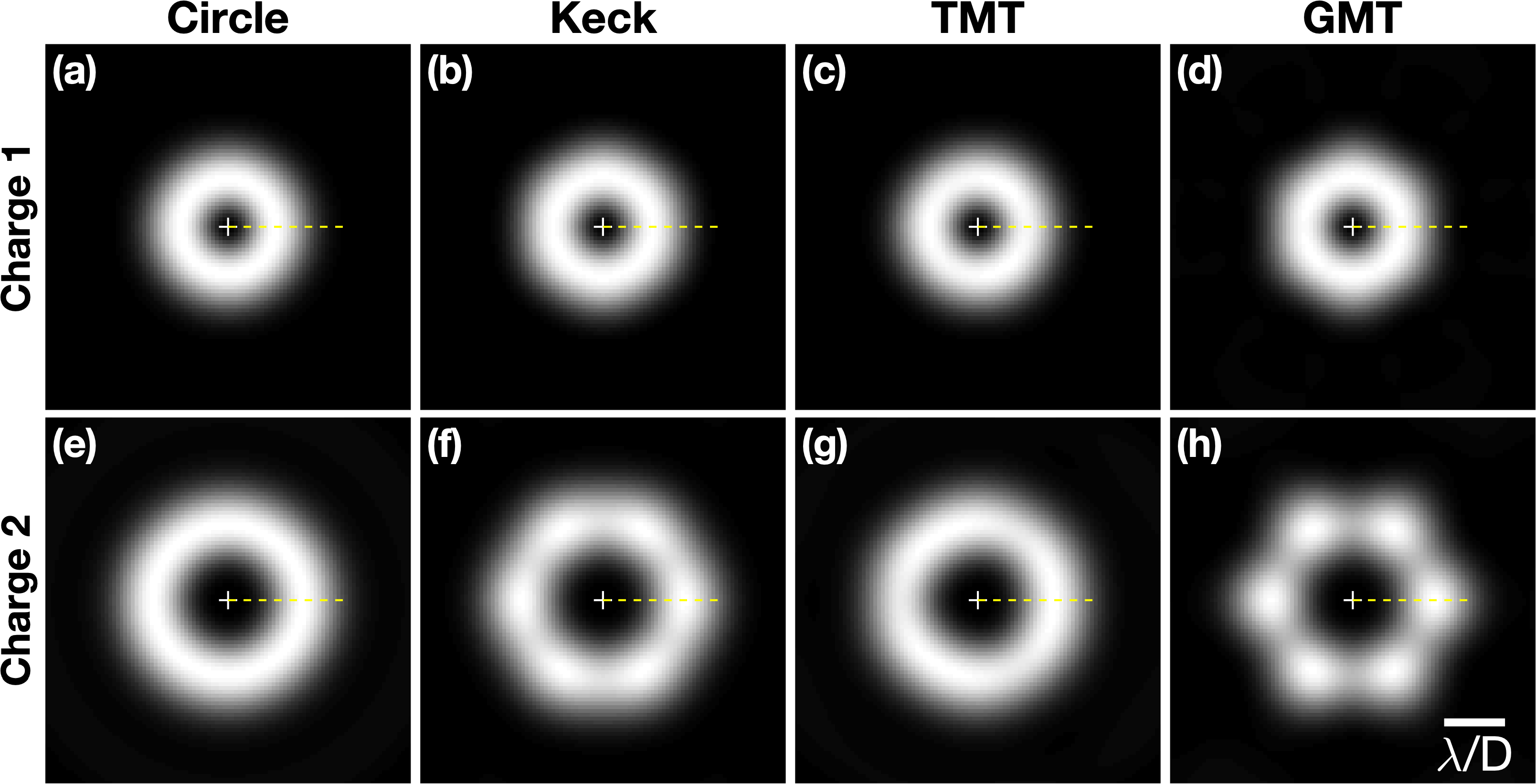}
    \caption{Maps of $\eta$ versus angular offset in two dimensions over a 6$\times$6~$\lambda/D$ square for each pupil (columns) and charge (rows). Figure~\ref{fig:throughput_grid} shows a horizontal line profile starting from the origin as indicated by the dashed yellow line. The throughput on non-circular apertures (e.g. GMT) has an weak modulation in the azimuthal direction. 
    }
    \label{fig:etamaps_pupils}
\end{figure}

Figure~\ref{fig:etamaps_pupils} shows maps of $\eta$ versus the two-dimensional position of the source, demonstrating that the fiber collects light from a donut shaped region around the star in all cases. This highlights a key advantage of the VFN approach: light from planets is collected regardless of the planet's position angle as projected on the sky, which is unknown for planets detected via the radial velocity or transit techniques. However, these techniques do constrain the separation of the planet from the star as a function of time, which is likely sufficient information to observe the planet with VFN. 

\begin{table}[t!]
    \centering
    \begin{tabular}{|c|c|c|c|c|c|c|c|}
        \hline
          & & \multicolumn{3}{c|}{Pupil plane VFN} & \multicolumn{3}{c|}{Focal plane VFN}  \\
        \hline
        Pupil & $l$ & $q_\mathrm{opt}$ & $\alpha_\mathrm{peak}$ ($\lambda/D$) & $\eta_\mathrm{peak}$ (\%) & $q_\mathrm{opt}$ & $\alpha_\mathrm{peak}$ ($\lambda/D$) & $\eta_\mathrm{peak}$ (\%)  \\
        \hline
         & 0 & 1.4 & 0.0 & 82 & - & - & - \\
        Circ & 1 & 1.4 & 0.85 & 19 & 2.5 & 0.84 & 20 \\
         & 2 & 1.3 & 1.3 & 10 & 3.6 & 1.3 & 10 \\
        \hline
         & 0 & 1.5 & 0.0 & 67 & - & - & - \\
        Keck & 1 & 1.4 & 0.90 & 18 & 2.6 & 0.90 & 19 \\
         & 2 & 1.4 & 1.5 & 11 & 3.8 & 1.5 & 11 \\
        \hline
         & 0 & 1.4 & 0.0 & 76 & - & - & - \\
        TMT & 1 & 1.4 & 0.87 & 19 & 2.5 & 0.87 & 20 \\
         & 2 & 1.3 & 1.4 & 9.6 & 3.6 & 1.4 & 9.9 \\
        \hline
         & 0 & 1.5 & 0.0 & 62 & - & - & - \\
        GMT & 1 & 1.4 & 0.88 & 15 & 2.6 & 0.87 & 16 \\
         & 2 & 1.3 & 1.5 & 10 & 3.6 & 1.4 & 10 \\
        \hline
    \end{tabular}
    \caption{Optimal parameters for pupil-plane and focal-plane VFN. $l$ is the charge of the vortex mask, $q_\mathrm{opt}$ is the value of $q=\mathrm{MFD}/(\lambda\,F\#)$ that maximizes the peak coupling efficiency (i.e. $\eta_\mathrm{peak}$), $\alpha_\mathrm{peak}$ is the angular separation at $\eta_\mathrm{peak}$, and $D$ is the circumscribed pupil diameter. }
    \label{tab:optparams}
\end{table}

\subsection{Layout considerations}

While the previous section demonstrated that there is little performance difference between pupil-plane and focal-plane layouts, there are several other pros and cons of each design that should be taken into consideration. 

\textbf{F\# compatibility.} An advantage of the pupil-plane layout is that the optimal focal ratio ($F\# = f/D$, where $f$ is the focal length) matches that of the standard FIU observing modes (stellar or direct exoplanet spectroscopy; Fig.~\ref{fig:obs_scenarios}a,b). Specifically, when coupling starlight in an SMF, the optimal value for the ratio $q=\mathrm{MFD}/(\lambda\,F\#)$, where MFD is the mode field diameter of the SMF, is $\sim$1.4. Table~\ref{tab:optparams} gives the optimal ratios, $q_\mathrm{opt}$, for each design simulated above. Whereas $q_\mathrm{opt}$ remains between 1.3 and 1.5 for the pupil-plane version for all aperture shapes and vortex charges, the focal plane version needs $q=2.5$ and $q=3.6$, for charge 1 and 2 VFN modes, respectively. Practically, this means the final focusing optics in the FIU would be the same for standard modes (Fig.~\ref{fig:obs_scenarios}a,b) and pupil-plane VFN (Fig.~\ref{fig:obs_scenarios}c), while focal-plane VFN would need lenses with different focal lengths. With a fixed MFD and $F\#$, the differences in $q_\text{opt}$ for each $l$ value in the pupil-plane case shifts the wavelength of maximal coupling a small amount, which tends to have a minor impact on the integrated coupling efficiency across typical astronomical passbands. 

\textbf{Beam and defect size.} Another advantage of the pupil-plane layout is that the central singularity of the vortex masks may be obscured by the shadow of the secondary mirror on ground-based telescopes. For instance, the image of the Keck pupil in the KPIC instrument is 12~mm and the central obscuration is 4~mm in diameter. By comparison, the focal plane version requires the central defect on the vortex mask to be on order of microns. On the other hand, since the beam at the vortex mask is $\sim$200$\times$ larger in the pupil-plane case, the transmitted wavefront error must be minimized over a larger area of the mask. 

\textbf{On-sky operations.} Pupil-plane VFN may also simplify on-sky observing procedures compared to focal-plane VFN. For pupil-plane VFN, the observer only needs to align the image of the star to the fiber, while in the focal-plane version the star must be accurately aligned to both the center of the vortex mask and the fiber.  

These practical trade-offs have led our team to adopt the pupil-plane version for current laboratory testing\cite{Echeverri2019_VFN} and future plans for on-sky deployment\cite{Echeverri2019b_VFN}, despite the fact that our first VFN paper\cite{Ruane2018_VFN} concentrated on the focal-plane approach. The remainder of this paper pertains to the pupil-plane version, but most of the theoretical performance properties also apply in the focal-plane case. 

\subsection{Sensitivity of the null depth to tip/tilt jitter and finite size of the star}

\begin{figure}[t!]
    \centering
    \includegraphics[height=0.32\linewidth]{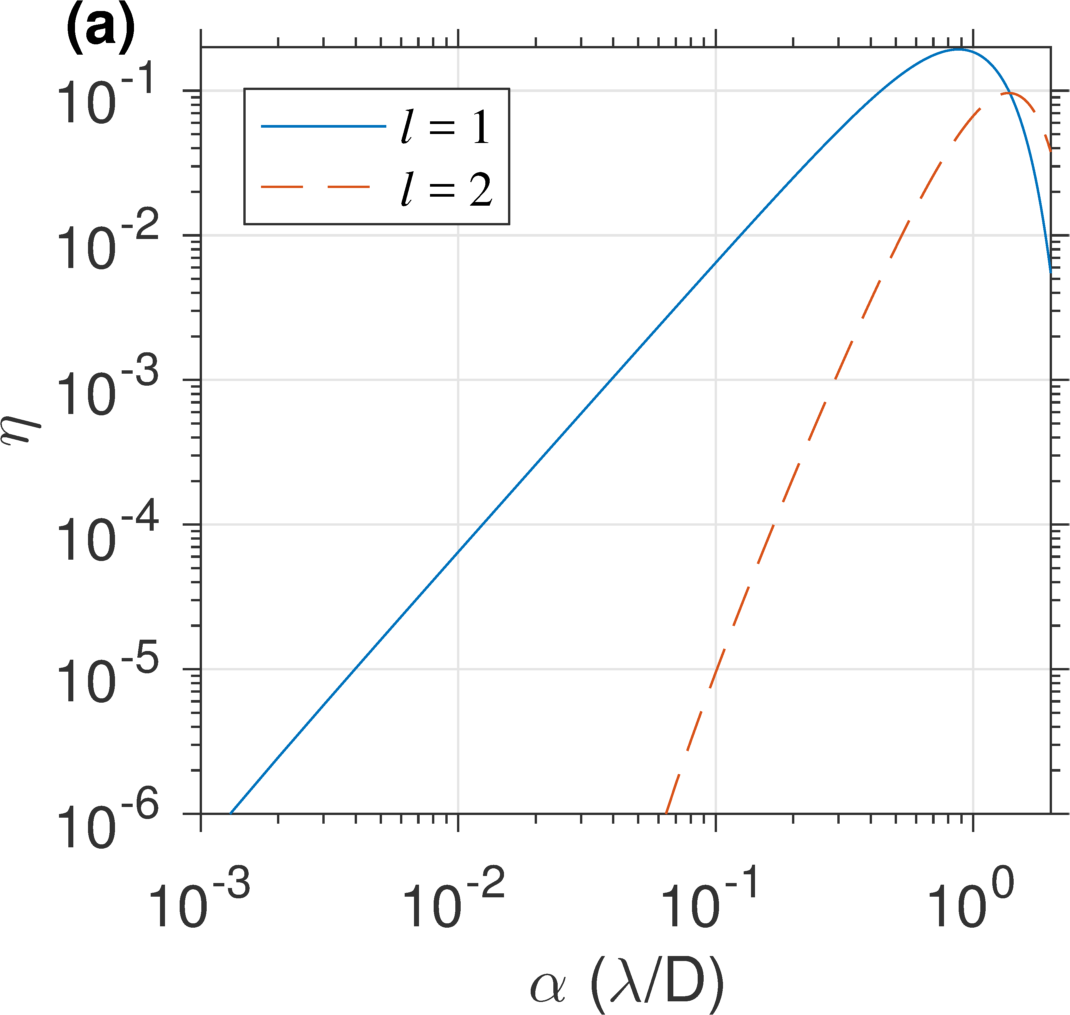}
    \includegraphics[trim={0.5cm 0 0 0},clip,height=0.32\linewidth]{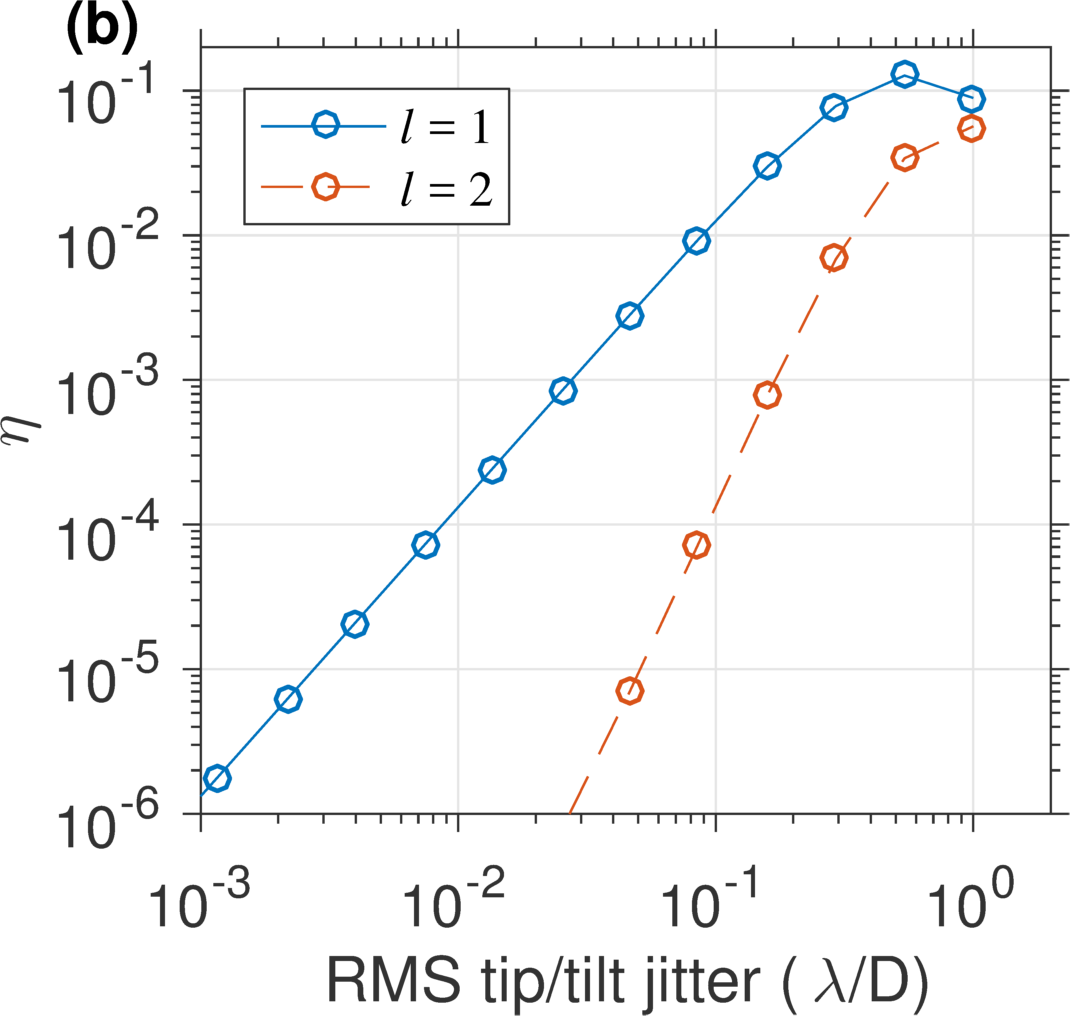}
    \includegraphics[trim={0.5cm 0 0 0},clip,height=0.32\linewidth]{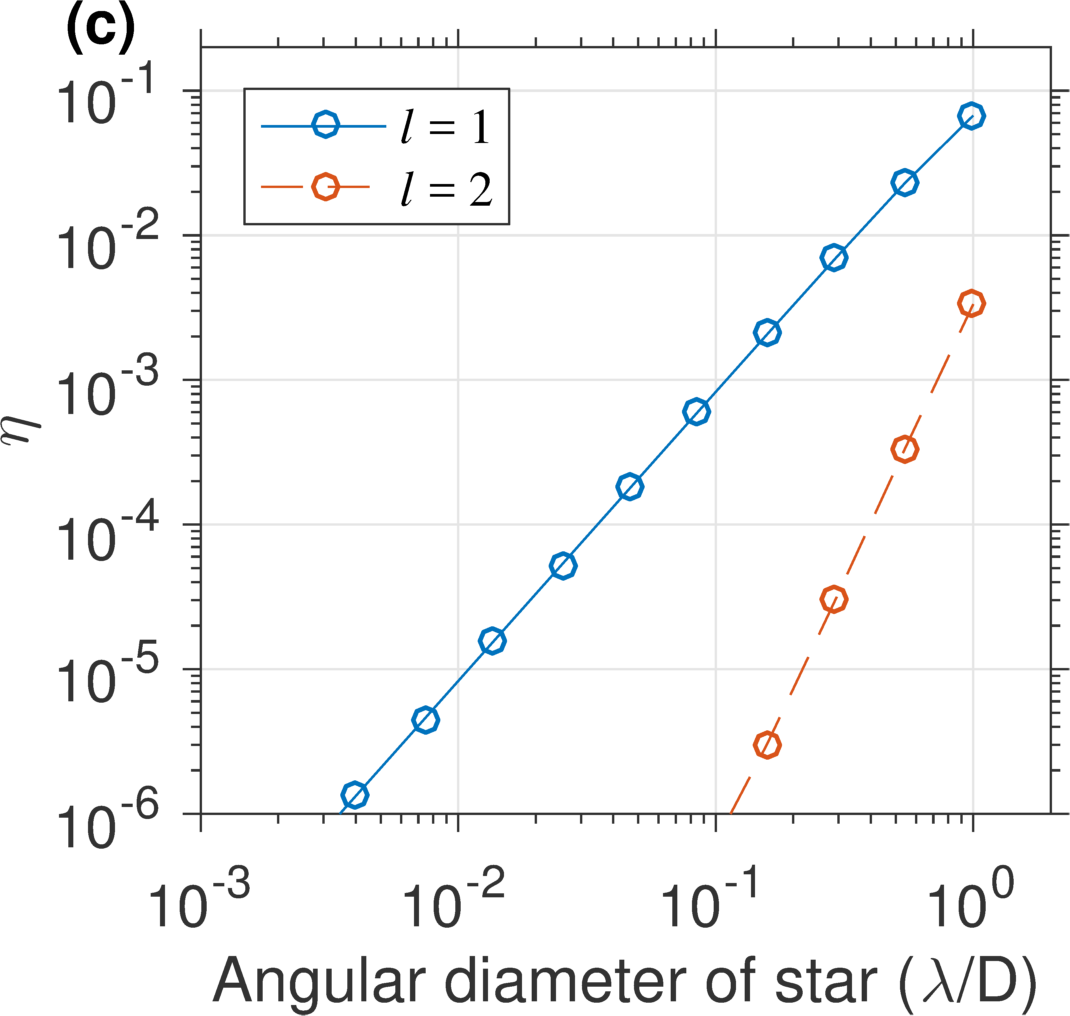} 
    \caption{Theoretical coupling efficiency of (a)~a point source as a function of angular offset from the optical axis and the fraction of starlight that couples into the fiber versus (b)~tip/tilt jitter and (c)~the size of the star. These values are calculated via a numerical simulations of the optical system with tip/tilt errors introduced in a Monte Carlo fashion. }
    \label{fig:ttsens}
\end{figure}

A deep null with VFN requires precise control of tip/tilt errors. Figure~\ref{fig:ttsens}a shows the coupling efficiency of a point source, $\eta$, versus its angular separation from the optical axis, $\alpha$, for charges $l=1$ and $l=2$. For $\alpha\ll\lambda/D$, the coupling efficiency follows a simple power law: $\eta\propto\alpha^{2l}$. Therefore, while reducing the charge provides better throughput for sources closer to the star, higher values of the charge are much more robust to small tip/tilt errors. Thus, the requirements for residual tip/tilt jitter after AO correction depends on the value of the charge (see Fig.~\ref{fig:ttsens}b). Assuming uncorrected errors are normally-distributed and time-averaged over a single exposure, the fraction of starlight that couples into the SMF (also refereed to here as the ``null depth") is approximately
\begin{equation}
    \eta_s\approx \left(\sigma \frac{D}{\lambda}\right)^{2l},
    \label{eqn:null_jitter}
\end{equation}
where $\sigma$ is the standard deviation of the tip/tilt error distribution in radians. For instance, to achieve $\eta_s=10^{-4}$, jitter needs to be less than 0.1~$\lambda/D$~RMS for $l=2$, which is routinely achieved by current ground-based AO systems. However, for the same null depth with $l=1$, the tip/tilt jitter requirement becomes roughly ten times smaller, which is challenging to achieve in the near-infrared with current instrumentation. 

To simulate a partially resolved star, we model it in a similar fashion, but as a disc of uniform emission. Figure~\ref{fig:ttsens}c shows the resulting null depth as a function of the angular size of the star with respect to the angular resolution of the telescope, $\lambda/D$. Although the $\eta$ values appear to be much smaller than the case shown for normally-distributed tip/tilt jitter, this is partially because the stellar size is specified by its full diameter rather than the standard deviation of the uniform distribution.

\subsection{Sensitivity of the null depth to low order aberrations}

\begin{figure}[t!]
    \centering
    \includegraphics[height=0.32\linewidth]{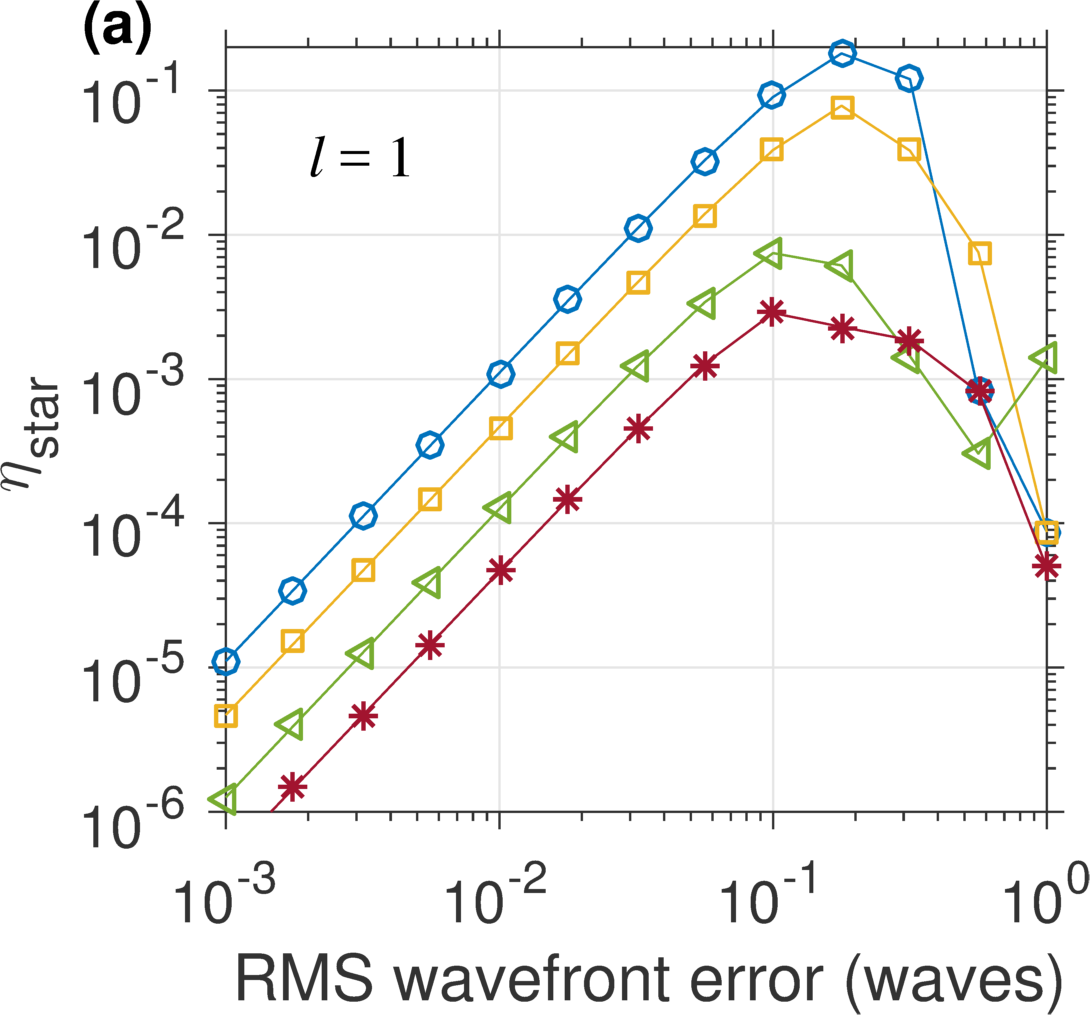}
    \includegraphics[trim={0.52cm 0 0 0},clip,height=0.32\linewidth]{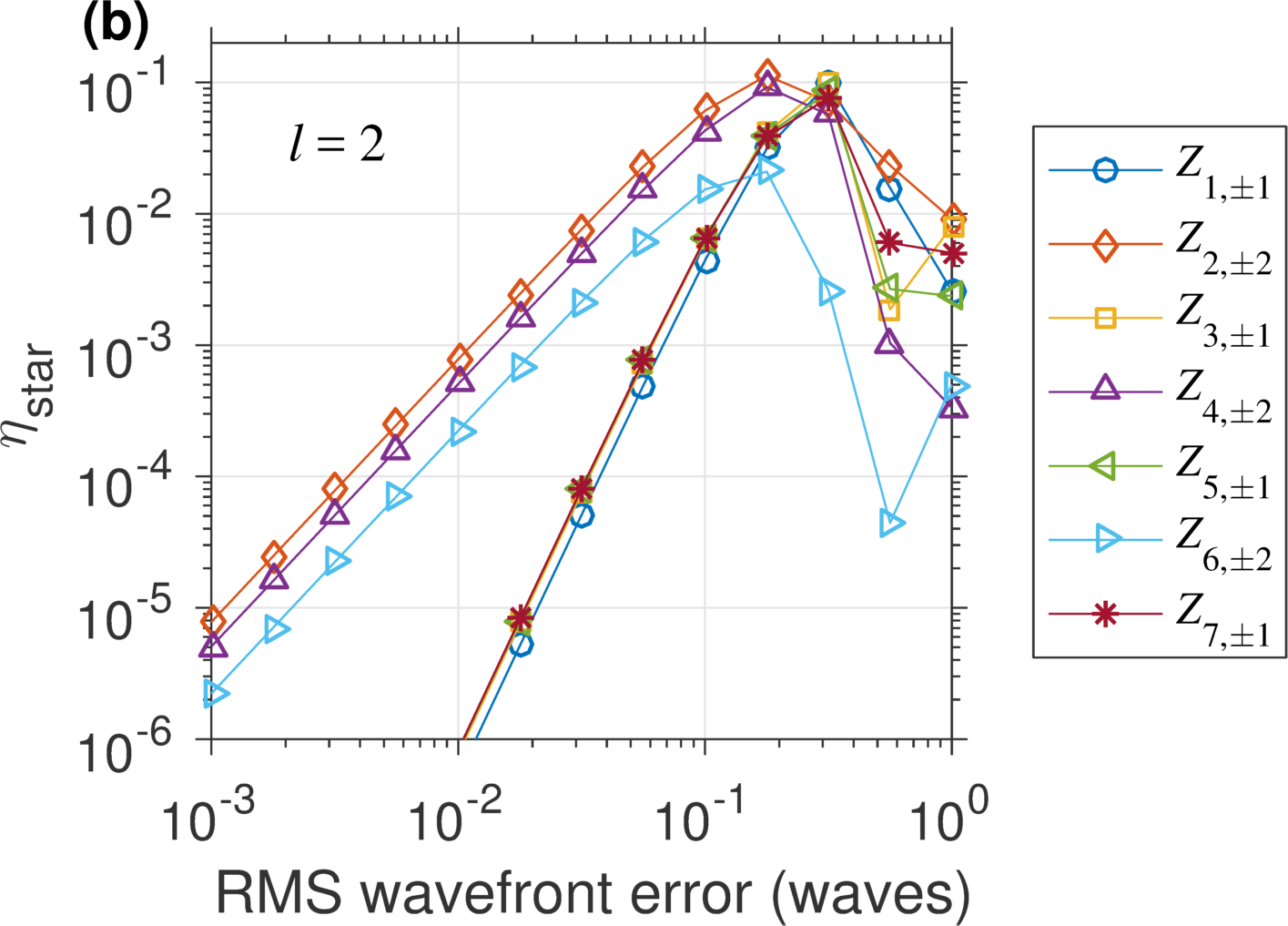}
    \caption{Sensitivity to low-order aberrations for charges (a)~$l=1$ and (b)~$l=2$ assuming the star is a point source and the pupil is circular. The modes that appear in (a) have azimuthal index $m=\pm1$ (i.e. tip, tilt, and all orders of coma) and the null depth follows second-order power law. The same modes appear in (b) with a fourth-order power law, while the null depth has a second-order sensitivity to modes with azimuthal index $m=\pm2$ (i.e. all orders of astigmatism). Terms with $m \ne \pm1$ or $m \ne \pm2$ are omitted because they do not contribute to the null depth in either case.}
    \label{fig:loworder}
\end{figure}

The performance of a VFN instrument is sensitive to some, but not all, low-order aberrations. Here, we describe the wavefront as a linear combination of Zernike polynomials. For small wavefront aberrations, $W(r,\theta)$, the field in the pupil is given by:
\begin{equation}
    E_p(r,\theta) = \exp(iW(r,\theta))\approx 1 + iW(r,\theta) = 1 + i\sum_{n,m}c_n^m R_n^m(r)\exp(im\theta),
    \label{eqn:loworderaber}
\end{equation}
where $c_n^m$ are coefficients and $R_n^m(\rho)$ are the radial Zernike polynomials\cite{BornWolf}. The field in the image plane due to $W(r,\theta)$ takes a similar form, but with a different set radial functions described by Bessel functions. Centering the vortex mask on the beam modifies the polar component associated with each term in the overlap integral to
\begin{equation}
    \int_0^{2\pi} \exp(il\theta)\exp(im\theta)d\theta = 
    \left\{
    \begin{array}{ll}
          2\pi & |l|=|m| \\
          0 & |l|\ne |m| \\
    \end{array} 
    \right..
    \label{eqn:nullcondition}
\end{equation}
Thus, the null depth will only be nonzero if $|m|=|l|$ and, under the first-order approximation, the charge~1 and~2 cases are most sensitive to coma ($m=\pm1$) and astigmatism ($m=\pm2$) orders, respectively. To illustrate this, Fig.~\ref{fig:loworder} shows the sensitivity of the null depth to each Zernike term. For the $l=1$ case, the null depth due to coma ($m=\pm1$) modes has second-order order dependence on wavefront error. For the $l=2$ case, coma ($m=\pm1$) modes have a 4th order dependence due to higher order terms in Eqn.~\ref{eqn:loworderaber}, while astigmatism ($m=\pm2$) terms have a second-order order dependence. In each case, the wavefront errors in only a few Zernike modes (where $|m|=|l|$) tend to limit the nulling performance.

\begin{figure}[t!]
    \centering
    \includegraphics[height=0.32\linewidth]{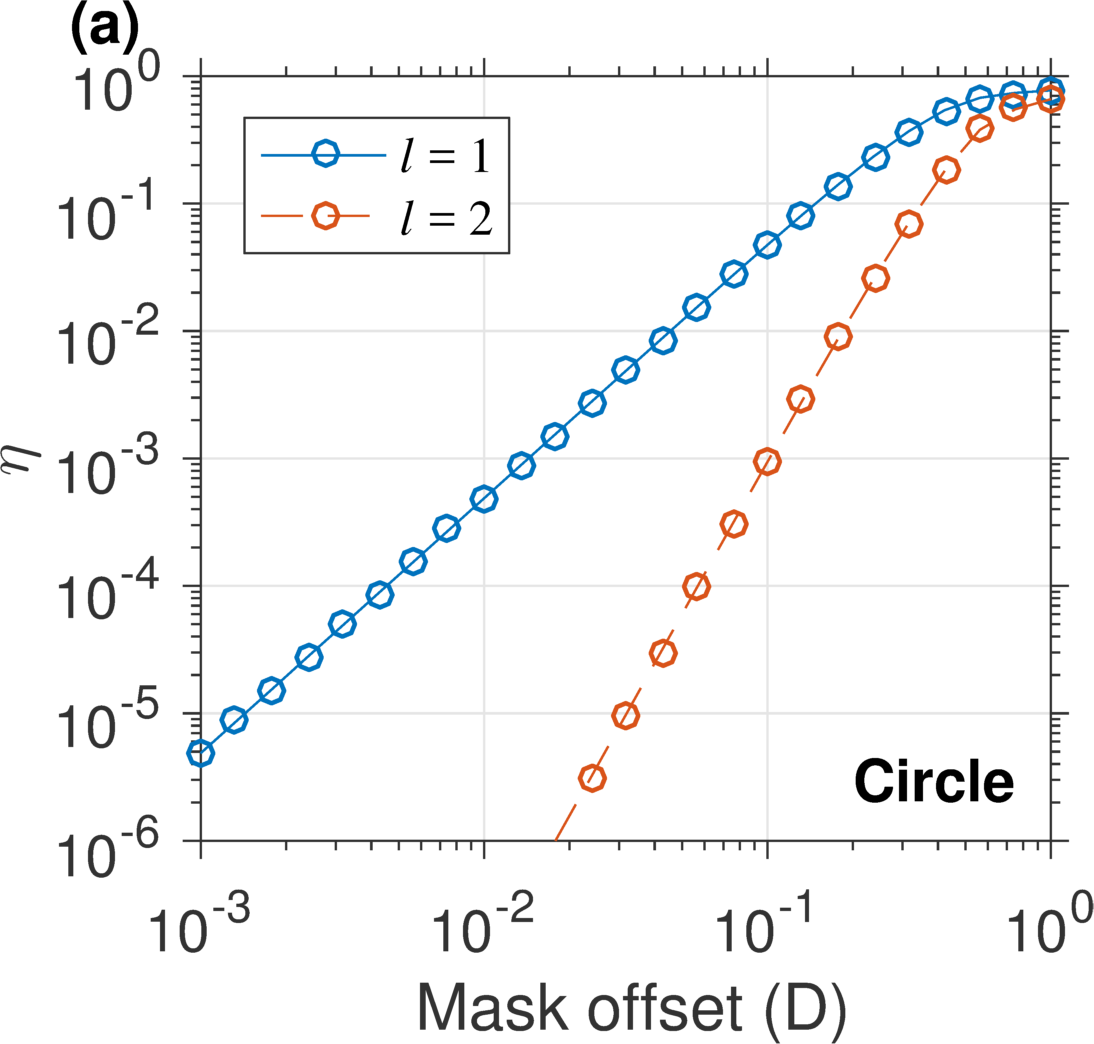}
    \includegraphics[trim={0.5cm 0 0 0},clip,height=0.32\linewidth]{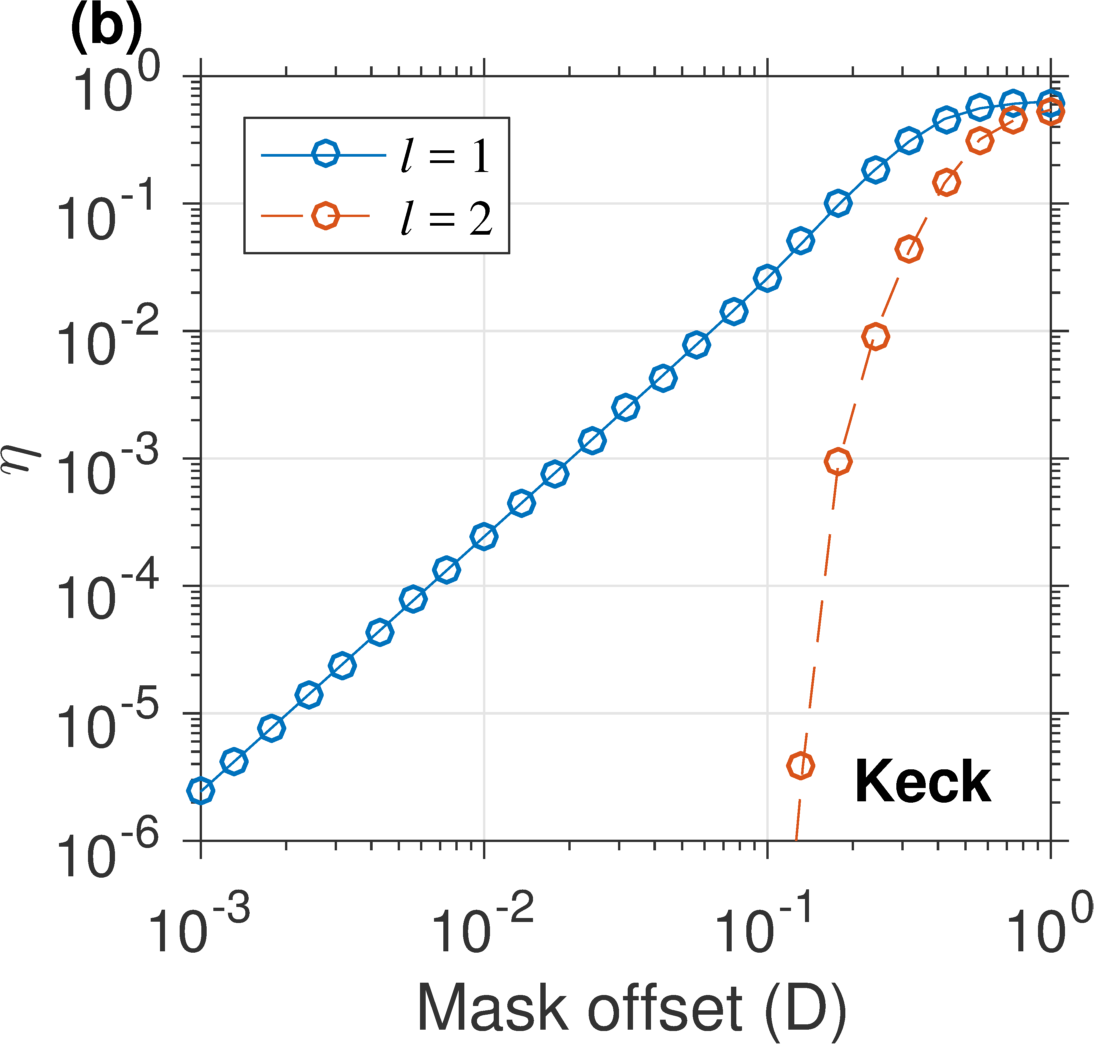}
    \includegraphics[trim={0.5cm 0 0 0},clip,height=0.32\linewidth]{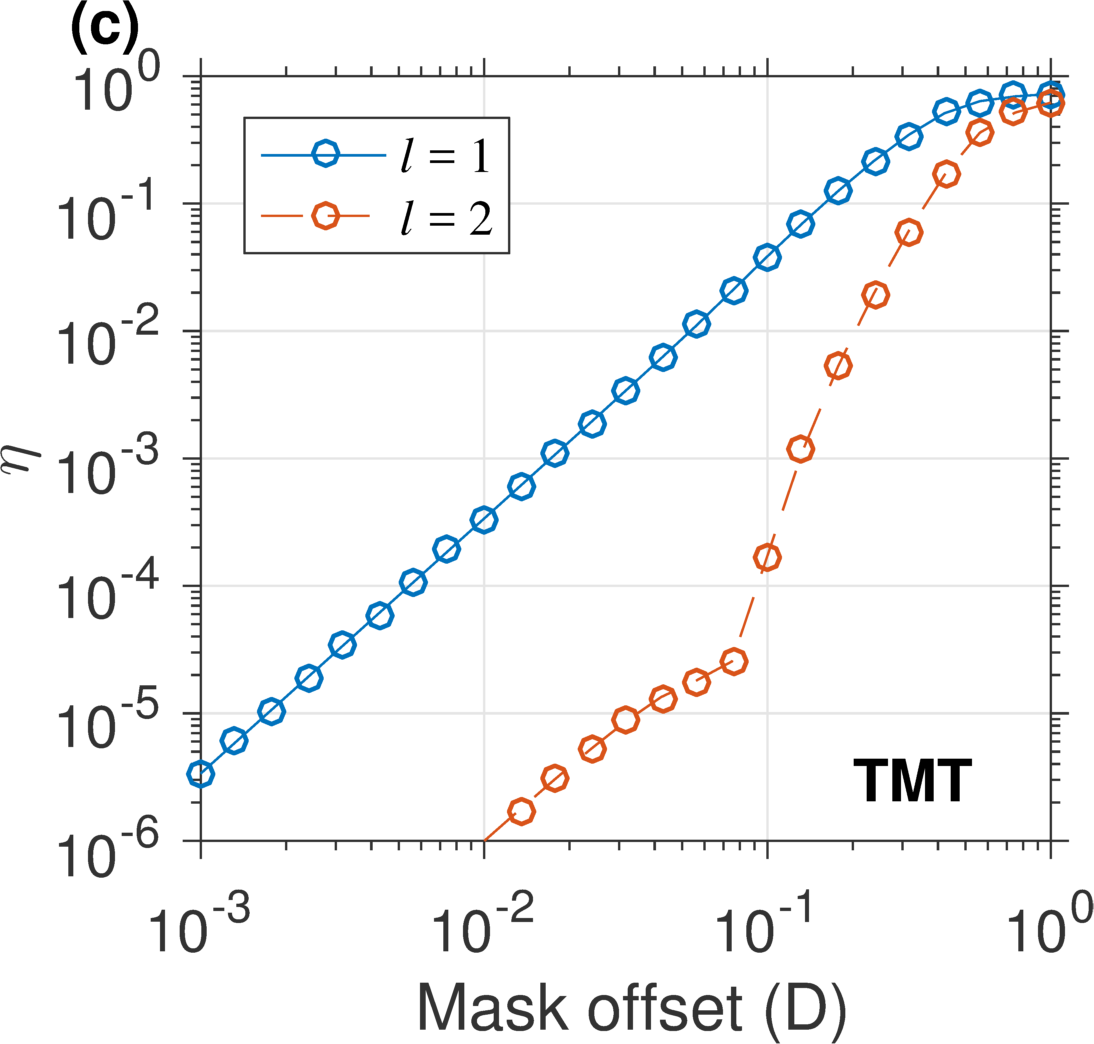} 
    \caption{Sensitivity to misalignment of the vortex mask for (a)~a circular pupil, (b)~the Keck telescope, and (c)~the TMT.}
    \label{fig:maskmisalignment}
\end{figure}

\subsection{Sensitivity to mask misalignment}

All of the previous calculations assumed that the vortex mask was perfectly centered on the pupil. However, the null depth degrades when the vortex mask is misaligned in the transverse direction. Figure~\ref{fig:maskmisalignment} shows how the null depth degrades as the mask is offset by a fraction of the pupil diameter for three pupil shapes. For $l=1$, the effect is similar for each pupil, where the null depth follows a well behaved second-order power law. However, for $l=2$, the null depth follows a fourth-order power law in the case of a circular aperture, but deviates from this behavior when the pupil is centrally obscured, as in the case of the Keck telescope, TMT, and GMT. We find that the null depth does not degrade significantly until the mask moves by approximately the radius of the central obscuration. For instance, for the Keck telescope, the diameter of the central obscuration is 24\% of the telescope outer diameter and, thus, a very low null depth is maintained until the mask is offset by $\sim$12\% of the pupil diameter (see Fig.~\ref{fig:maskmisalignment}b).

\subsection{Mask designs}

There are several varieties of vortex masks in the literature; most types fall into two categories: \textit{vectorial} and \textit{scalar} masks. 

\subsubsection{Vector vortex fiber nuller}

Vector phase masks may be manufactured via a variety of methods: liquid crystal (LC) \cite{Marrucci2006,Mawet2009,Mawet2010a,Mawet2010b,Serabyn2019}, subwavelength gratings\cite{Biener2002,Mawet2005b,Niv2007}, and photonic crystals\cite{Murakami2013}. Each impart conjugate phase patterns to the incident circular polarization states, while also inverting the two states. The phase pattern is set by the local orientation of the fast axis, i.e. the phase shift is $\Phi=\pm 2\chi(x,y)$, where $\chi(x,y)$ is the fast-axis angle. Thus, a vector vortex mask has $\chi=l\theta/2$ and Jones matrix in the circular polarization basis, $\mathbf{M}_\circlearrowright$, which acts on an incoming beam as follows:
\begin{equation}
\left[ \begin{matrix}
   U^\prime_R \left(x,y\right) \\
   U^\prime_L \left(x,y\right)  \\
\end{matrix} \right]=\mathbf{M}_\circlearrowright \left(x,y\right)  \left[ \begin{matrix}
   U_R \left(x,y\right)  \\
   U_L \left(x,y\right)  \\
\end{matrix} \right]
=\left[\begin{matrix}
   0 & e^{il\theta}  \\
   e^{-il\theta} & 0  \\
\end{matrix} \right] \left[ \begin{matrix}
   U_R \left(x,y\right)  \\
   U_L \left(x,y\right)  \\
\end{matrix} \right],
\end{equation}
where $U_R \left(x,y\right)$ and $U_L \left(x,y\right)$ are the right- and left-handed circularly polarized field components in the $(x,y)$ plane. However, the half-wave condition is not achieved at all wavelengths simultaneously, which causes an additional leakage term, which may be represented by:
\begin{equation}
\mathbf{M}_\circlearrowright 
=c_V \left[\begin{matrix}
   0 & e^{il\theta}  \\
   e^{-il\theta} & 0  \\
\end{matrix} \right]
+ c_L \left[\begin{matrix}
   1 & 0  \\
   0 & 1  \\
\end{matrix} \right],
\end{equation}
where $c_V$ and $c_L$ are wavelength-dependent coefficients, $|c_L|^2\approx\epsilon(\lambda)^2/4$, and $\epsilon(\lambda)$ is the retardance error\cite{Mawet2010a}. The null depth is given by $\eta_s=|c_L|^2\eta_0$, where $|c_L|^2$ is fraction of light leaked through the vector vortex with $l=0$ and $\eta_0$ is the coupling efficiency for the $l=0$ mode (see Table \ref{tab:optparams}). For $|c_L|^2<10^{-4}$, the retardance error requirement is $\sim1^\circ$, which may be achieved over substantial bandwidths using multi-layer LC\cite{Pancharatnam1955,Komanduri2013}. 

\subsubsection{Scalar vortex fiber nuller}

Scalar vortex approaches using spiral phase plates\cite{Swartzlander2008,Ruane2019_scalarVC} or dispersion-compensated holograms\cite{Errmann2013,Ruane2014}, though appearing earlier in the literature\cite{Heckenberg1992,Tidwell1993,Beijersbergen1994}, have not received as much attention in the context of exoplanet detection and characterization methods. However, there are several viable methods for implementing a scalar vortex fiber nuller.  

Perhaps the simplest scalar vortex mask is a spiral phase plate, which has transmittance of the form $t(\lambda)=\exp\left(il(\lambda)\theta\right)$, where $l(\lambda)$ is the effective charge as a function of wavelength, which takes on non-integer values:
\begin{equation}
    l(\lambda) = l_0\frac{\lambda_0}{\lambda}\left(\frac{n(\lambda)-1}{n(\lambda_0)-1}\right),
\end{equation}
where $l_0$ is the the charge at the central wavelength $\lambda_0$ and $n(\lambda)$ is the refractive index of the material. A spiral phase plate with a wavelength-independent refractive index or a spiral phase mirror\cite{Campbell2012} would have a wavelength-dependent charge profile given by $l(\lambda)=l_0\lambda_0/\lambda$. For the sake of simplicity, we simulate the spiral phase mirror case below. 

\begin{figure}[t!]
    \centering
    \includegraphics[height=0.32\linewidth]{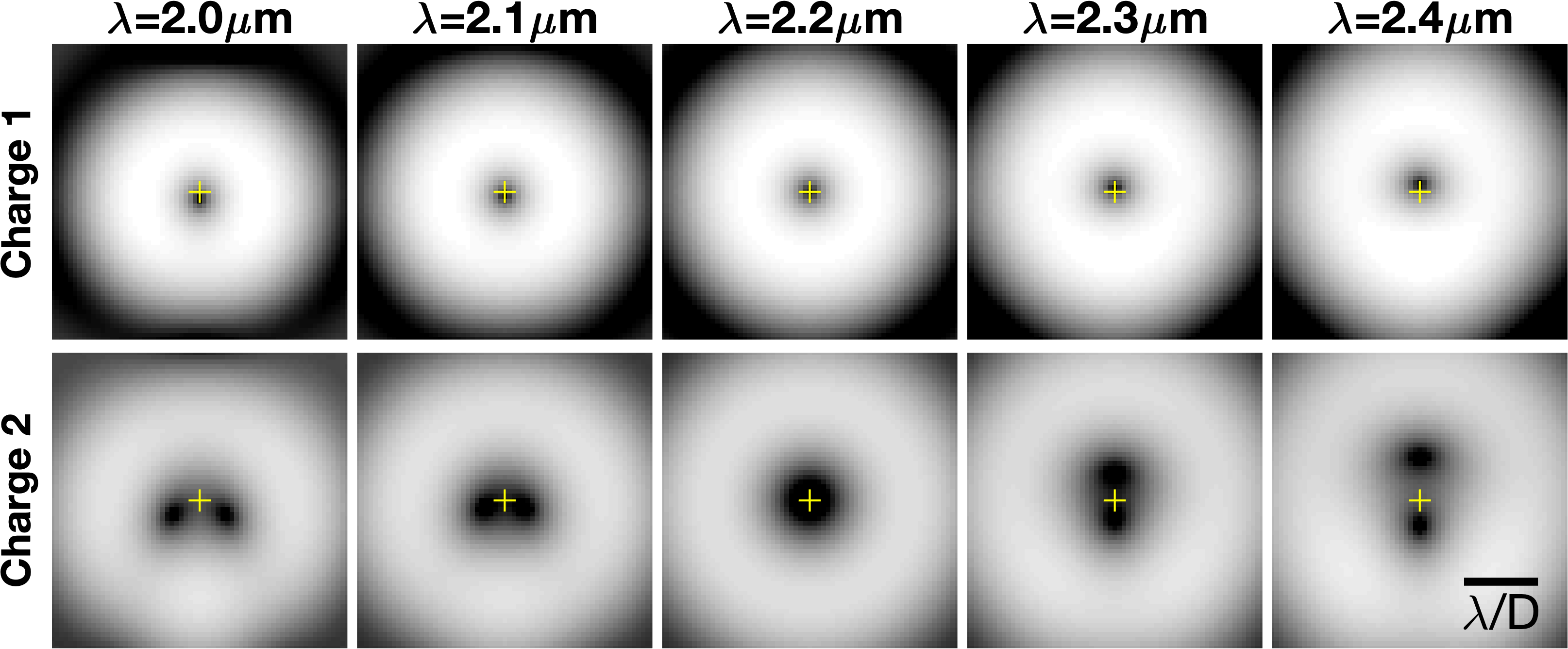}
    \caption{Two-dimensional maps of $\log(\eta)$ versus angular separation for a scalar VFN for (top row) $l_0=1$ and (bottom row) $l_0=2$ at $\lambda_0=2.2~\mu m$. By comparison, a vector vortex is designed to give charge $\pm l_0$ at all wavelengths. }
    \label{fig:etamap_scalar_vs_wvl}
\end{figure}

\begin{figure}[t!]
    \centering
    \includegraphics[height=0.32\linewidth]{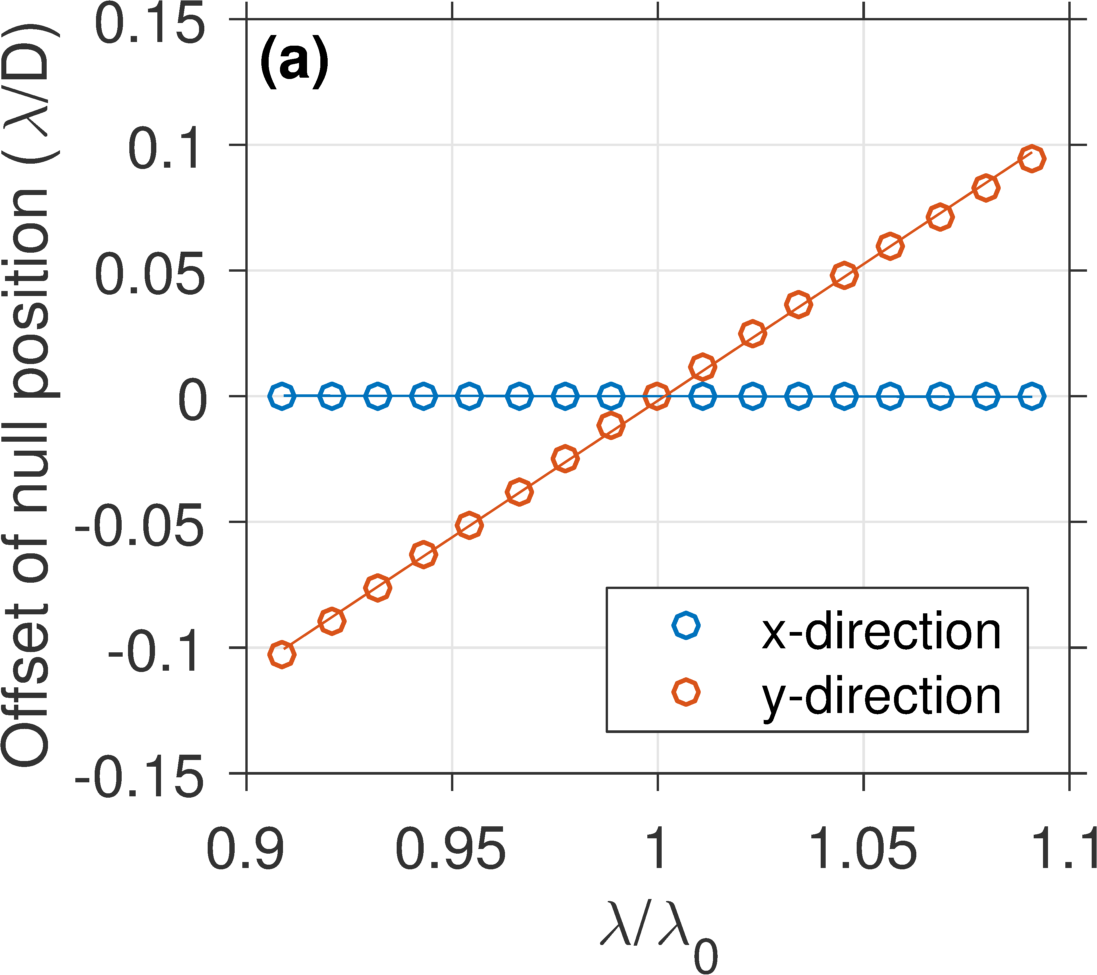}
    \includegraphics[height=0.32\linewidth]{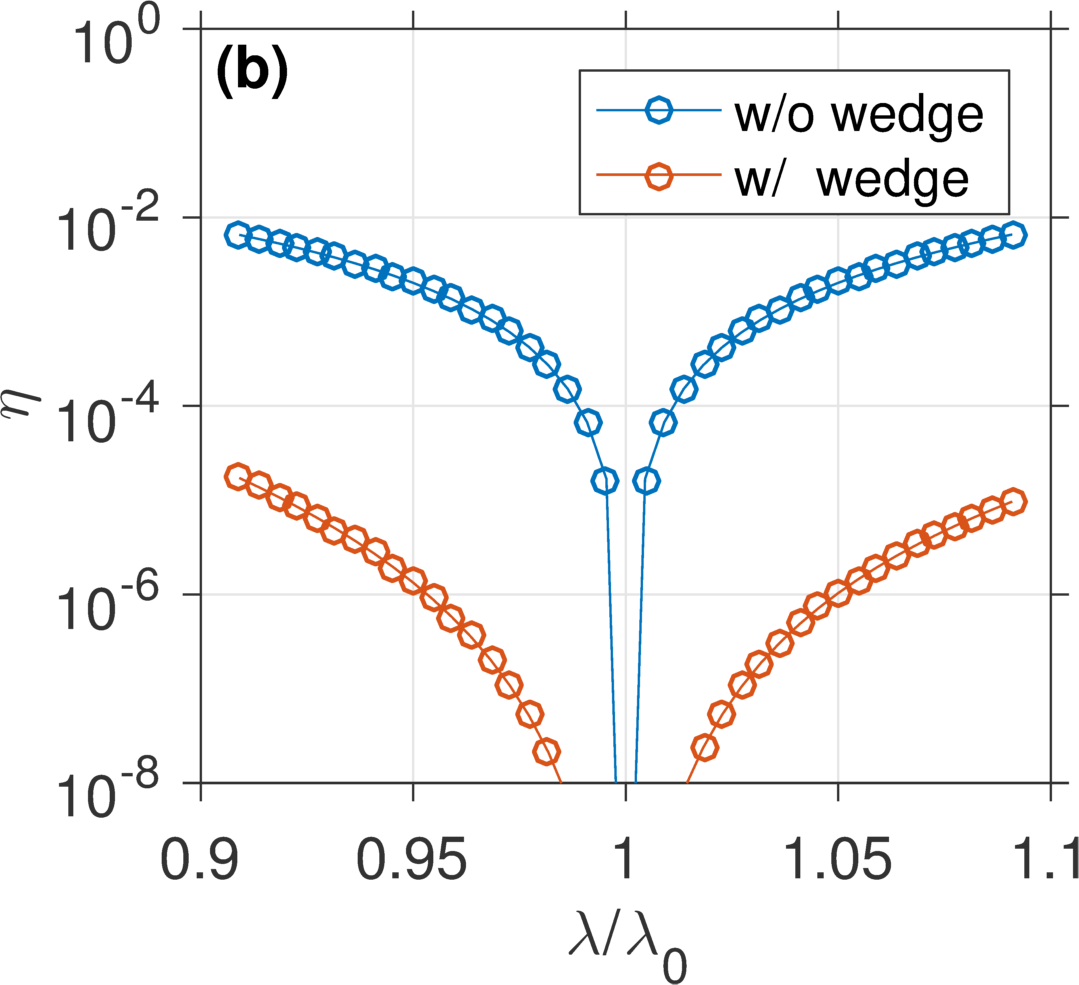}
    \caption{Correction of the wavelength-dependent offsets of an $l=1$ scalar VFN by adding a wedge of dielectric. (a)~The offsets seen in the $\eta$-maps in the top row of Fig.~\ref{fig:etamap_scalar_vs_wvl}. (b)~Comparison of null depth before and after adding a wedge with $n=1.4$ and $d\tan(\beta)=6~\mu m$.}
    \label{fig:scalaroffsets}
\end{figure}

Figure~\ref{fig:etamap_scalar_vs_wvl} shows maps of $\eta$ versus angular offset for a scalar VFN using a spiral phase mirror with charge $l_0$ at a central wavelength of 2.2~$\mu m$. Although a true null only occurs at the central wavelength for a fiber centered on the ``yellow cross," a null exists at other wavelengths with a wavelength-dependent offset from the origin. Therefore, a null may be created at any given wavelength by shifting the image of the star with respect to the fiber by the equal and opposite amount. In the charge 1 case, the null appears to follow a straight line as a function of wavelength, whereas in the charge 2 case, the null breaks up into two charge 1 nulls that move in opposite directions. At least in the charge 1 case, the wavelength-dependent offset of the null (see Fig.~\ref{fig:scalaroffsets}a) can be mitigated to first order by inserting a wedge of dielectric material. Since a peak-to-valley phase shift, $\Delta\Phi$, of one wave corresponds to a $\lambda/D$ offset in the image plane, the wedge must apply
\begin{equation}
    \Delta\Phi = 2 \pi p \left( \frac{\lambda-\lambda_0}{\lambda} \right) + c,
\end{equation}
where $p=1.1$ (i.e. the slope of the linear fit in Fig.~\ref{fig:scalaroffsets}a) and $c$ is a constant representing a global offset in the image plane. A prism with refractive index $n$ and wedge angle $\beta$, applies 
\begin{equation}
    \Delta\Phi = \frac{2 \pi}{\lambda}(n-1)d\tan(\beta),
\end{equation}
where $d$ is the diameter of the beam. We find the wavelength-independent solution by setting $c=2\pi p$ (i.e. the broadband null will occur $p\;\lambda_0/D$ from the optical axis) and manufacture the prism such that
\begin{equation}
    d\tan(\beta) = \frac{p\;\lambda_0}{n-1}.
\end{equation}
Figure~\ref{fig:scalaroffsets}b shows $>$300$\times$ improvement in the null depth after introducing a prism with $n=1.4$ and $d\tan(\beta)=6~\mu m$. In this example, the phase discontinuity occurs along the negative $x$-axis in the pupil, the offset moves along the positive $y$-axis with increasing wavelength, and therefore the wedge must compensate for dispersion in the $y$-direction (i.e. perpendicular to the phase discontinuity).

Our team is actively exploring alternate approaches for implementing scalar VFN, including imprinting the vortex pattern directly on the deformable mirror, segmented primary mirror\cite{Tyson2008}, or the tip of the SMF\cite{Vayalamkuzhi2016}. All of these methods are viable as long as the wavelength-dependent tilt can be accurately removed by the prism, or similar optic. In practice, the atmospheric dispersion compensator (ADC) may take the place of the prism, thereby potentially eliminating the need to add any new optic to existing AO-fed FIU instruments in order to implement VFN. Future work will explore the trade-offs between such methods.

\subsection{Beam shaping: a potential pathway towards improving planet coupling efficiency}

\begin{figure}[t!]
    \centering
    \includegraphics[height=0.33\linewidth]{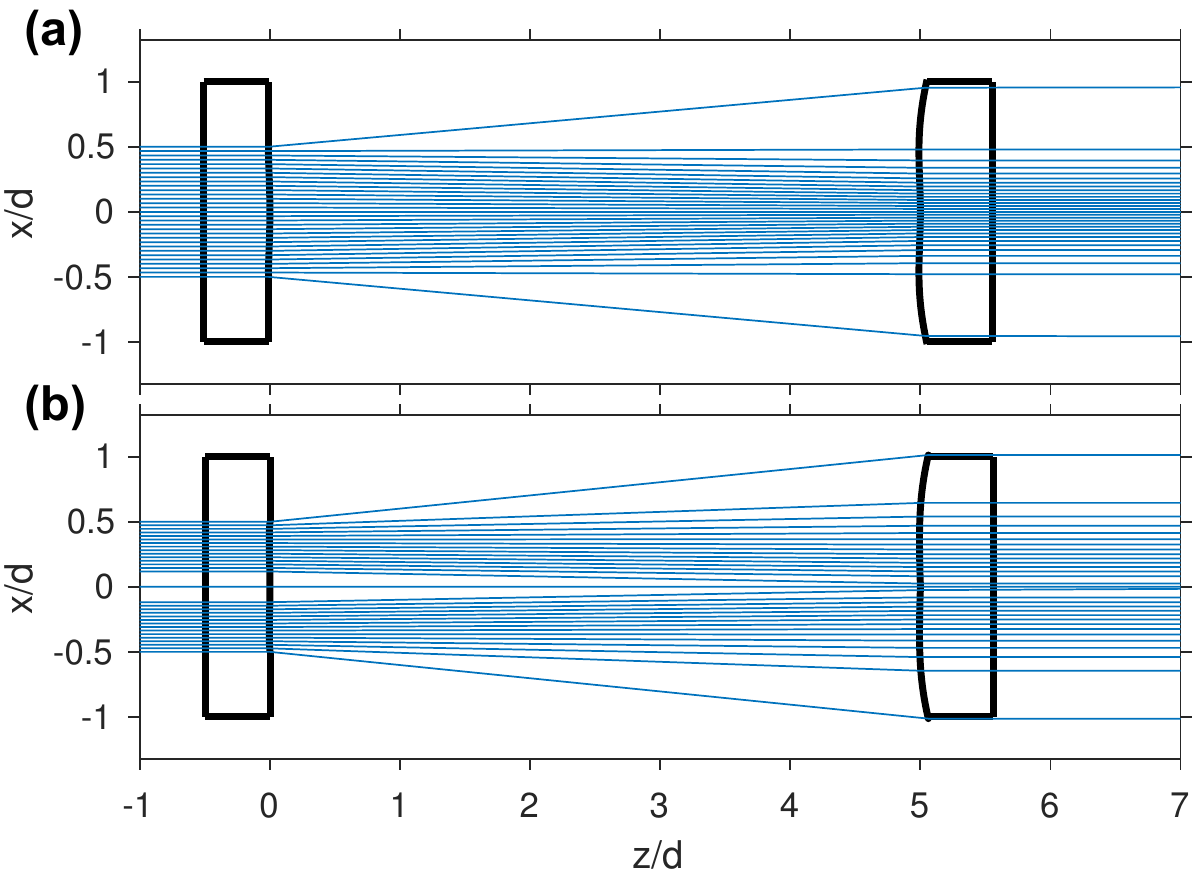}
    \includegraphics[height=0.32\linewidth]{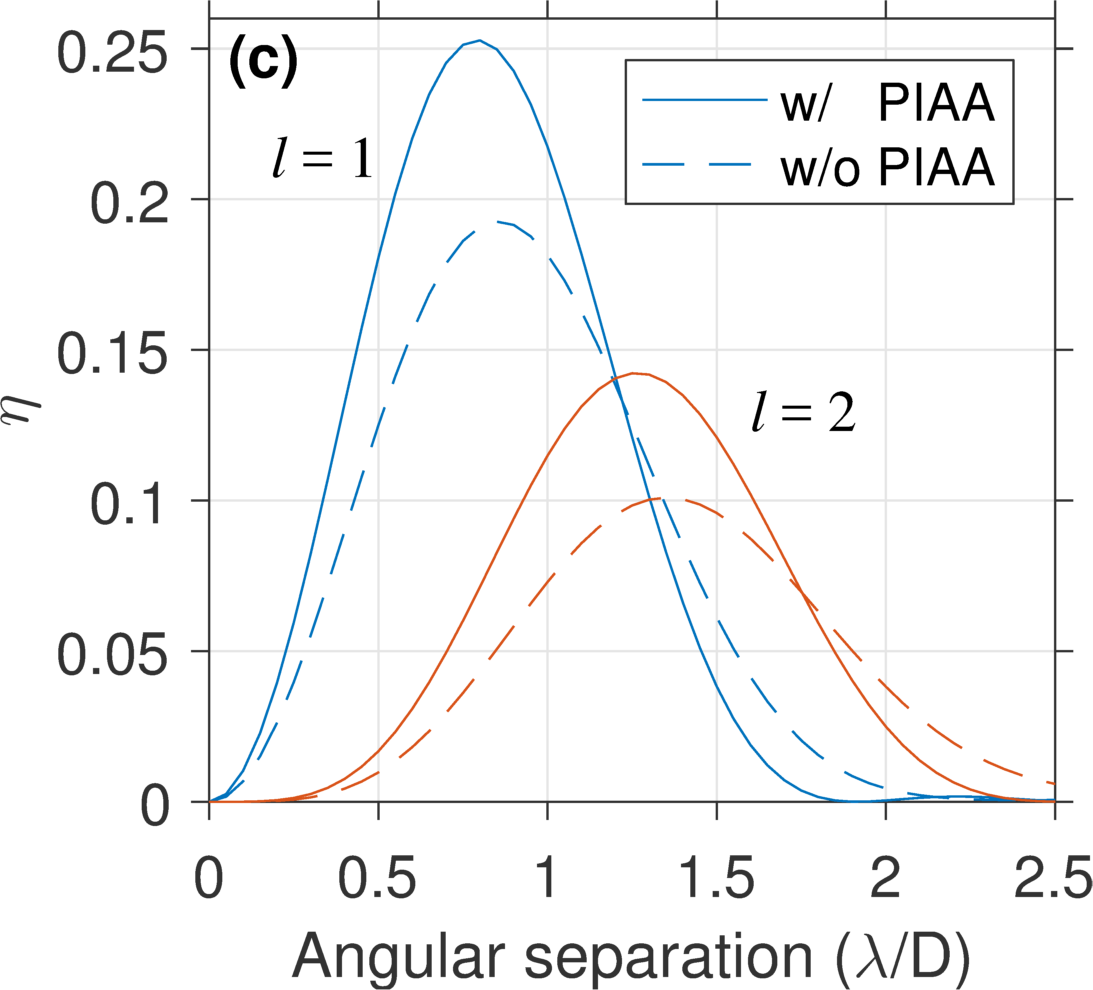}
    \caption{Beam shaping with phase-induced amplitude-apodization (PIAA) lenses. (a)-(b)~Ray trace through PIAA lenses designed for (a)~a circular beam and (b)~an annular beam. (c)~The coupling efficiency as a function of angular separation of the planet from the optical axis, $\alpha$, (solid lines)~with and (dashed lines)~without the PIAA lenses as well as (blue lines)~$l=1$ and (red lines)~$l=2$ VFN modes. The coupling efficiencies shown are for a circular pupil, but we achieve similar gains for a large range of telescope pupils.}
    \label{fig:piaa}
\end{figure}

We are also exploring methods to improve the VFN throughput. Beam shaping optics have been previously demonstrated to improve coupling efficiency of light from a large telescope into SMFs\cite{Jovanovic2017}. Building upon this work, we designed phase-induced amplitude-apodization (PIAA) lenses\cite{Guyon2005} to remap the collimated beam into a quasi-Gaussian beam in order to increase the coupling efficiency for the planet light in a VFN instrument. Figure~\ref{fig:piaa} shows two example PIAA lens pairs that are designed for circular (Fig.~\ref{fig:piaa}a) and annular (Fig.~\ref{fig:piaa}b) input beams. The latter is based on the dimensions of the Keck pupil. While each are designed to maximize the coupling efficiency in the $l=0$ case, they also provide significant throughput improvement in $l=1$ and $l=2$ VFN modes (see Fig.~\ref{fig:piaa}c). Specifically, for a circular pupil, the peak coupling efficiency increases by 20\%, 30\%, and 40\% in the $l=0$, $l=1$, and $l=2$ cases, respectively, while for the Keck pupil, the coupling increases by approximately 30\% for all three cases. 

Using PIAA lenses for beam shaping along with VFN may not require additional engineering in practice since PIAA lenses are already likely to be a key component of AO-fed FIUs for stellar spectroscopy and direct exoplanet spectroscopy where the target of interest is aligned to the optical axis (as in Fig.~\ref{fig:obs_scenarios}). A potential drawback of the PIAA lenses is that they have an extremely narrow field of view (1-2~$\lambda/D$) within which wavefront aberrations are small enough to allow high coupling efficiency and, therefore, they are not applicable for simultaneous spectroscopy of multiple objects. In that case, image slicing methods may be necessary to achieve high efficiency using a pair of PIAA lenses and SMF for each object. 

\section{THE POTENTIAL SCIENTIFIC YIELD FOR VFN ON TMT}

The capability to probe very small angular separations may naturally lead to high scientific yield because the occurrence rate of planets is potentially much greater for closer-in orbits than direct imagers are currently able to observe (10-100~au). In fact, RV and direct imaging surveys suggest that there may be a peak in the occurrence rate of giant planets within 1-10~au\cite{Nielsen2019}, which may be accessible in the near-infrared with a nuller on TMT beyond distances of 100~pc. VFN can therefore be used to detect and characterize young (1-10~Myr) giant planets in several nearby star forming regions and young moving groups\cite{Bowler2016}. VFN can also target smaller planets whose occurrence rate increases to almost one per star for periods less than 50 days and radii 0.5-4~$R_\oplus$ \cite{Dressing2013}. Though, the latter will lead to considerably longer integration time, $\tau$, which scales as $r_p^{-4}$ when observing a planet of radius $r_p$ in reflected light\cite{Ruane2018_VFN}.

\begin{figure}[t]
    \centering
    \includegraphics[height=0.33\linewidth]{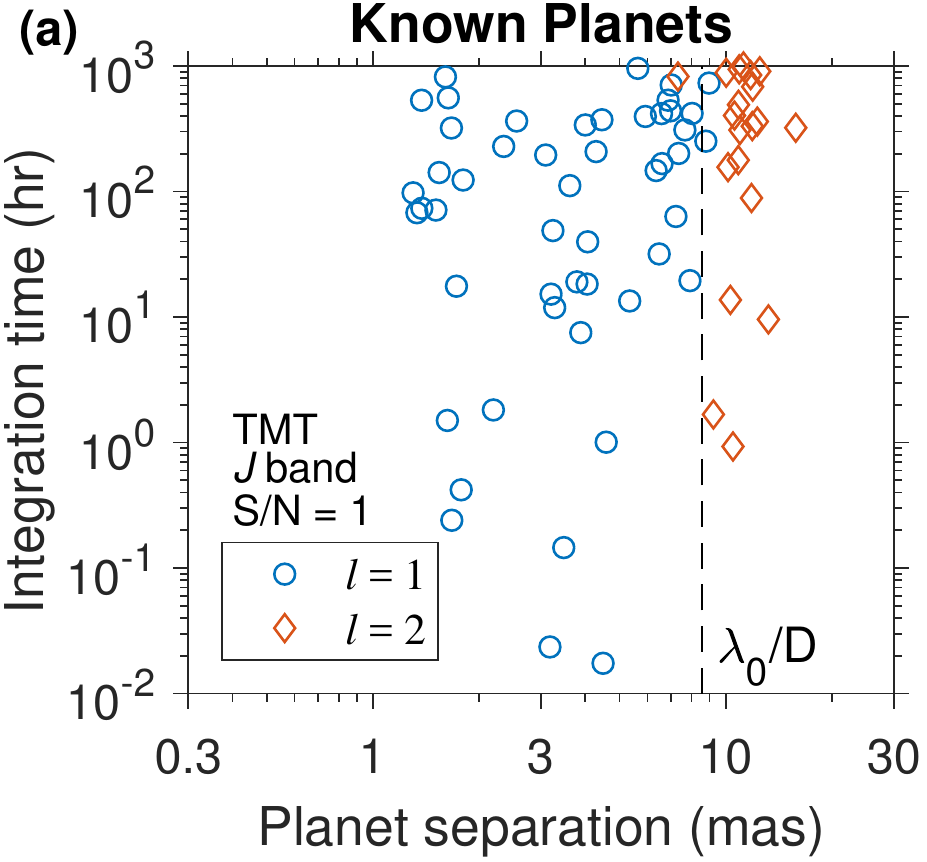}
    \includegraphics[height=0.33\linewidth]{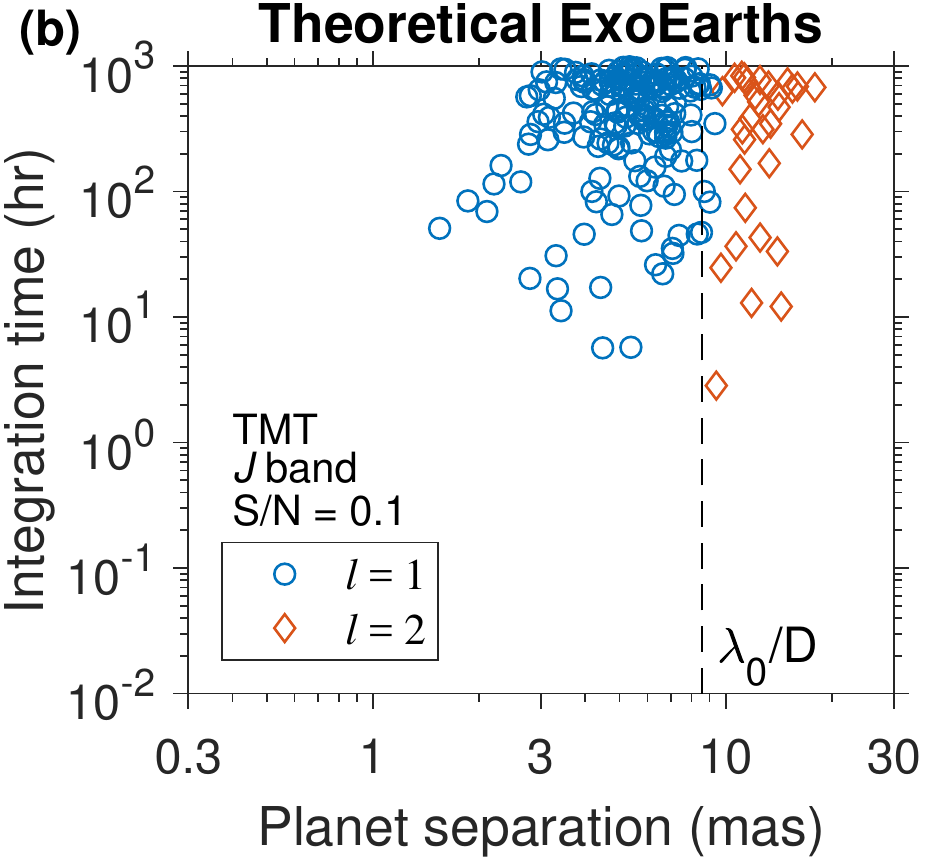}
    \caption{Integration time estimates for achieving the goal $S/N$ per spectral channel at a spectral resolution of $R$=100,000 in $J$ band for (a)~known exoplanets and (b)~theoretical exoEarths orbiting all nearby stars. We set the goal $S/N$ per spectral channel to 1 and 0.1 in each case, respectively. The blue circles indicate cases where $l=1$ provides a shorter integration time and likewise the red diamonds are cases where $l=2$ is more favorable. In this scenario, $l=2$ tends to be beneficial when the planets angular separation from the host star is $>\lambda_0/D$=8.6~mas, where $\lambda_0=1.25~\mu$m.}
    \label{fig:tints}
\end{figure}

Possibly the most immediate application of VFN on TMT will be to independently confirm the existence of previously detected planets using other techniques, such as the RV or transit methods. To determine the potential number of feasible targets, we computed the $\tau$ needed detect all of the planets listed in the NASA Exoplanet Archive (accessed Dec 2018) at $S/N=1$ per spectral channel. To compute the planet-to-star flux ratio, we used the open-source atmospheric modeling package \texttt{PICASO}\cite{Batalha2019} to generate both the reflected light and thermal emission of the planet, assuming a single layer cloud deck in the atmosphere. Our current cloud model yielded an albedo of $\sim~0.2$ in $J$-band, but we note that the choice of cloud model can change reflected light flux by factors of several. The stellar size is calculated using an empirical radius-magnitude relationship for late-type stars\cite{Mann2015}. The telescope and instrument are assumed to have collecting area of 655~m$^2$, spectral resolution $R= 100,000$, transmission of 30\%, detector quantum efficiency of 90\%, no other detector noise, RMS tip-tilt jitter = $10^{-2}~\lambda/D$, an ideal vector vortex mask, and no PIAA optics. These calculations only take into account photon noise due to leaked starlight (i.e the dominant noise source) with the appropriate VFN throughput and expected leakage due to the estimated stellar size in addition to the assumed tip/tilt jitter. For each target, we determine the value of $l$ that gives the shortest integration times for these assumptions. Figure~\ref{fig:tints}a shows the integration times in $J$ band. The number of targets for which $S/N=1$ is reached in $\tau<$50~hr is 23, 17, and 23 in $J$, $H$, and $K$ bands, respectively. The targets with the shortest theoretical integration times are \textit{tau~Boo~b} and \textit{ups~And~b}. While the calculated $\tau$ for most planets is lower for $l=1$ under these assumptions, some cases prefer the $l=2$ mask, which are mostly scenarios where the planet separation is $>\lambda/D$; an example of such a case is \textit{55~Cnc~b} in $J$ band, which requires $\tau=2$~hr. A number of planets with relatively short integration time are within the conventional definition of the inner working angle (roughly 0.4~$\lambda/D$), demonstrating that VFN is beneficial even at separations of relatively low throughput (i.e. less than half of its maximum). 

VFN may also be a pathway towards detecting and characterizing Earth-sized planets in the habitable zone around nearby, cool stars. Figure~\ref{fig:tints}b shows the estimated number of such stars around which $S/N=0.1$ could be achieved on a theoretical 1~$R_\oplus$ planet at the center of the habitable zone (the input catalog is freely available at Olivier Guyon's webpage\cite{guyonwebpage}). We otherwise apply equivalent assumptions as listed above. The number of targets where $\tau<$50~hr is 23, 18, and 10 in $J$, $H$, and $K$ bands, respectively. Since the prime targets tend to be nearby stars, the stellar angular sizes are larger with respect to $\lambda/D$ and therefore $l=2$ is more often favorable for this science case. One of the best targets is \textit{Wolf~359}, which can be detected at $J$, $H$, and $K$ bands with integration times of $\sim$3~hr. While using $l=2$ in $J$ band leads to a shorter $\tau$ for \textit{Wolf~359}, $l=1$ would be preferred for $H$ and $K$ bands where the planet-star separation is too small for $l=2$. 




\section{CONCLUSIONS}

VFN is a promising method for detecting and characterizing exoplanets with large-aperture ground- and space-based telescopes. Here, we showed that the small inner working angle provides access to close-in planets with separations from their host star on the order of $\lambda/D$. Increasing the charge of the vortex mask provides a way to reduce sensitivity to tip/tilt and mask misalignments at the expense of increasing the inner working angle. We demonstrated that (a)~the theoretical performance of VFN is largely insensitive to the shape of the telescope aperture and some low order aberrations, (b)~both vector and scalar implementations are possible, and (c)~beam shaping can improve the throughput. Moreover, we showed that VFN is not sensitive to the planet's position angle, which implies that the RV or transit methods provide sufficient information for follow up observations using VFN. We expect to be able to detect $>$10 known planets with integration times on the order of tens of hours using VFN on an AO-fed FIU on TMT, such as the MODHIS instrument\cite{Mawet2019_whitepaper}, and possibly even enable the detection Earth-sized planets in the habitable zone of cool stars.

\acknowledgments  
This work was carried out at the Jet Propulsion Laboratory, California Institute of Technology, under contract with the National Aeronautics and Space Administration (NASA). This research has made use of the NASA Exoplanet Archive, which is operated by the California Institute of Technology, under contract with NASA under the Exoplanet Exploration Program.


\small
\bibliography{Library}   
\bibliographystyle{spiebib}   

\end{document}